\documentclass{article}
\usepackage[preprint]{neurips_2026}
\usepackage[utf8]{inputenc}
\usepackage[T1]{fontenc}
\usepackage{hyperref}
\usepackage{url}
\usepackage{booktabs}
\usepackage{amsfonts}
\usepackage{amsmath}
\usepackage{amssymb}
\usepackage{amsthm}
\usepackage{nicefrac}
\usepackage{microtype}
\usepackage{graphicx}
\usepackage{natbib}
\usepackage{algorithm}
\usepackage{algorithmic}
\usepackage{adjustbox}
\usepackage{tabularx}
\usepackage{enumitem}
\usepackage{wrapfig}
\usepackage{placeins}
\usepackage[table]{xcolor}
\usepackage[most]{tcolorbox}
\tcbuselibrary{skins,breakable,listings}
\definecolor{APSTableBlue}{RGB}{235,242,255}
\definecolor{APSTableGreen}{RGB}{238,247,242}
\definecolor{APSTableLavender}{RGB}{246,242,252}
\newcommand{\validationdash}{%
  \noalign{%
    \vskip 0.4pt%
    \begingroup
    \color{gray!60}%
    \hbox to \textwidth{%
      \xleaders\hbox{\rule{7pt}{0.65pt}\hspace{2.5pt}}\hfill\kern0pt%
    }%
    \endgroup
    \vskip 0.4pt%
  }%
}

\newcommand{\APScaptionfont}{\fontsize{9pt}{10pt}\selectfont}

\tcbset{
  guiguardprompt/.style={
    enhanced jigsaw,
    breakable,
    colback=orange!7!white,
    colbacklower=orange!7!white,
    colframe=orange!75!black,
    colbacktitle=orange!80!black,
    coltitle=white,
    interior style={fill=orange!7!white},
    frame style={draw=orange!75!black},
    fonttitle=\bfseries,
    title=Prompt,
    boxrule=0.8pt,
    arc=2pt,
    left=8pt,
    right=8pt,
    top=7pt,
    bottom=7pt
  }
}
\newtcblisting{promptbox}[1][]{
  guiguardprompt,
  listing only,
  listing options={basicstyle=\ttfamily\small,breaklines=true,columns=fullflexible},
  #1
}

\makeatletter
\long\def\@makecaption#1#2{%
  \vskip\abovecaptionskip
  \sbox\@tempboxa{\APScaptionfont #1. #2}%
  \ifdim \wd\@tempboxa >\hsize
    \APScaptionfont #1. #2\par
  \else
    \global \@minipagefalse
    \hb@xt@\hsize{\hfil\box\@tempboxa\hfil}%
  \fi
  \vskip\belowcaptionskip}
\newcommand{\AppendixContentsStart}{%
  \let\APS@oldsection\section
  \let\APS@oldsubsection\subsection
  \renewcommand{\section}[1]{%
    \APS@oldsection{##1}%
    \addcontentsline{apc}{section}{\protect\numberline{\thesection}##1}%
  }%
  \renewcommand{\subsection}[1]{%
    \APS@oldsubsection{##1}%
    \addcontentsline{apc}{subsection}{\protect\numberline{\thesubsection}##1}%
  }%
}
\newcommand{\AppendixContentsPage}{%
  \clearpage
  \begin{center}
    {\Large\bfseries Appendix\par}
  \end{center}
  \vspace{1.2em}
  \noindent{\Large\bfseries Contents}\par
  \vspace{0.25em}
  \noindent\rule{\textwidth}{0.4pt}\par
  \vspace{0.25em}
  \begingroup
    \setcounter{tocdepth}{2}%
    \renewcommand*\l@section[2]{%
      \@dottedtocline{1}{0pt}{2.4em}{\bfseries ##1}{\bfseries ##2}%
    }%
    \renewcommand*\l@subsection[2]{%
      \@dottedtocline{2}{2.4em}{3.0em}{##1}{##2}%
    }%
    \@starttoc{apc}%
  \endgroup
  \vspace{0.4em}
  \noindent\rule{\textwidth}{0.4pt}
  \clearpage
  \AppendixContentsStart
}

\newcommand{\AppendixContentsStop}{%
  \let\section\APS@oldsection
  \let\subsection\APS@oldsubsection
}

\newcommand{\sidecaptionof}[2]{\par\vspace{3pt}\refstepcounter{#1}\noindent{\APScaptionfont\parbox[t]{\linewidth}{\APScaptionfont\normalfont\noindent\csname fnum@#1\endcsname. #2\par}}\par}
\makeatother

\title{APS: Bias-Controlled Adaptive Prototype Simulation for Population-Scale LLM Agents}

\author{%
\parbox{0.90\textwidth}{\centering
\textbf{Quan Zheng}$^{1,2}$, 
\textbf{Yan Gao}$^{2,3}$\thanks{Corresponding author.}, 
\textbf{Shaobin He}$^{2}$, 
\textbf{Haoxiang Guan}$^{2}$\\
\textbf{Yuanhe Tian}$^{2,3}$, 
\textbf{Jie Feng}$^{2,3}$, 
\textbf{Ming Wang}$^{1}$, 
\textbf{Shuxin Zheng}$^{2,3}$, 
\textbf{Zhen Liu}$^{2}$\\[0.6ex]
$^{1}$Beijing Normal University\\
$^{2}$Zhongguancun Academy, Beijing, China\\
$^{3}$Zhongguancun Institute of Artificial Intelligence\\[0.5ex]
{\small\texttt{s-zq25@bza.edu.cn, gaoyan@zgci.ac.cn}}
}%
}

\begin{document}
\maketitle

\begin{abstract}
LLM-agent simulation offers a flexible computational tool for studying population response trajectories that depend on scenario events, memory, demographics, and evolving social context. However, full multi-round simulation scales linearly with both population size and horizon, requiring every agent to query the LLM at every round. We propose \textbf{Adaptive Prototype Simulation (APS)}, a framework that reframes scalable LLM-based simulation as a recurrent oracle-allocation problem. APS retains the designated LLM as the online transition oracle while querying adaptive core prototypes, selected
singleton-tail agents, and shadow-audit agents. Prototype responses induce local
response surfaces for nearby agents, reducing online LLM calls without replacing
the underlying transition model. To control approximation bias, shadow-audit residual correction estimates propagation residuals for aggregate correction and future budget allocation, while tail-protected singleton routing directly queries selected isolated, heterogeneous, or high-curvature regions that are vulnerable to smoothing. Theoretically, we treat APS as an estimator for full-scale high-precision individual social simulation, and decompose its errors into  prototype-coverage error, shadow-audit residual-correction error, local-propagation bias, and temporal context mismatch. Under the reported protocols, APS gives lower reference-aligned distributional discrepancy than scale-oriented and same-budget baselines while reducing online LLM calls, with ablations and compact robustness checks diagnosing the main bias-control mechanisms. In a 10M-agent, multi-round public-opinion simulation, APS achieves a $381.1\times$ reduction over full simulation, with reference-aligned final-round JSD 0.094 against the corresponding full-LLM reference.
\end{abstract}

\section{Introduction}

Large language models make it possible to build social agents whose responses depend on scenario prompts, memory, demographics, and evolving social context \citep{park2023generative,wang2024surveyagents,guo2024multiagent}. This can expand classical agent-based modeling beyond hand-crafted transition rules \citep{epstein1996growing,macy2002factors,castellano2009statistical}, but it creates a computational bottleneck: simulating $N$ agents for $T$ rounds requires $O(NT)$ online LLM calls.

Existing strategies either keep rich prompts but simulate only small ``micro-societies'' \citep{park2023generative,li2023camel}, or reduce repeated LLM calls through serving optimizations, routing, or learned proxies \citep{frantar2023gptq,leviathan2023fast,chen2023frugalgpt,kwon2023efficient,guan2025earthscale}. These lines reduce different costs, but they do not by themselves provide a recurrent population-transition estimator that keeps the specified LLM in the online loop while accounting for propagation bias, tail smoothing, and temporal context mismatch.

We propose \textbf{Adaptive Prototype Simulation (APS)}, a framework for scalable LLM-agent population simulation that treats scaling as recurrent LLM-oracle allocation. APS selects representative prototypes, queries their LLM-induced responses under the current scenario and social context, and applies these transitions to similar agents. Across rounds, APS reallocates prototype budgets using stratum size, residual error, and local similarity and curvature diagnostics rather than a fixed representative schedule. Compared with prior cost-reduction strategies for large-scale simulation, APS reduces online calls without replacing the specified LLM transition with a trained proxy; queried prototype LLM responses remain the semantic anchors for same-round propagation.

The key challenge is approximation bias. Agents assigned to similar prototypes can still exhibit different LLM-induced transitions, and this mismatch can affect later prompts through recurrent states and neighbor summaries. APS uses two complementary bias-control mechanisms. Shadow-audit residual correction uses shadow-audit queries to estimate discrepancies between local-response-surface predictions and individual LLM responses; these residuals correct aggregate transition estimates and inform future budget allocation toward high-error, high-curvature, or low-recall regions. Tail-protected singleton routing directly queries selected isolated feature-space agents and removes them from the smoothing operator, reducing the risk that selected isolated, heterogeneous, or high-curvature feature-space regions are absorbed by dominant prototypes.

We study the computational problem of approximating the population transition induced by full LLM-agent simulation, not the behavioral-validity problem of proving that LLMs model human populations. The central idea is to turn prototype propagation from an unchecked compression heuristic into an audited oracle-allocation loop: prototypes anchor same-round semantics, tail-protected singleton routing reduces smoothing of selected isolated agents, and shadow-audit residuals correct aggregate estimates while steering future queries. APS is specified by oracle anchoring, tail-protected singleton routing, shadow-audit residual correction, and the separation of recurrent hard states from audit-corrected reporting; the partitioner and local-response-surface operator are replaceable implementation choices, and our empirical claims are limited to the reported protocols.

Our contributions are organized in three layers:
\begin{itemize}
    \item \textbf{APS framework for population-scale LLM agents.} APS keeps the LLM as the online transition oracle, queries adaptive core prototypes each round, and propagates their same-round transitions to the full population while maintaining recurrent social context. Its allocation rule dynamically shifts prototype budgets across strata according to residual error, local similarity diagnostics, and stratum size, rather than fixed representative quotas.
    \item \textbf{Bias control for prototype propagation.} APS reduces approximation bias introduced by prototype propagation through shadow-audit residual correction and tail-protected singleton routing. Shadow-audit residuals correct aggregate estimates and inform future budget allocation toward high-error, high-curvature, or low-recall regions; tail-protected singleton routing directly queries selected isolated agents vulnerable to prototype smoothing.
    \item \textbf{Analysis and large-scale evaluation.} We formalize the dynamic LLM transition estimand, decompose error into prototype coverage, shadow-audit residual-correction, core local-propagation, and temporal context mismatch terms, and evaluate APS through a 10M-agent, 8-round simulation with $381.1\times$ fewer calls, same-budget baselines, ablations, a 10K trajectory check, and scenario, LLM backend, and stratification backend robustness checks.
\end{itemize}

\vspace{-0.1cm}

\section{Related Work}

Generative agents and related systems show that LLMs can drive memory-based routines, social interaction, role play, and survey-like behavior in small-to-medium populations \citep{park2023generative,li2023camel,chen2024agentverse,zhou2024sotopia,argyle2023out,aher2023using,park2024generative1000,ashery2025emergent}. As researchers use LLM agents to study population-level regularities closer to real-world social dynamics, recent surveys document this shift, and large-scale systems and the large population models framework scale simulation via platforms, learned surrogates, differentiable population-scale frameworks, and topology-aware grouping \citep{gao2024llmabm,mou2024individual,gao2024s3,vezhnevets2023concordia,tang2024gensim,piao2025agentsociety,guan2025earthscale,chopra2025largepopulationmodels,xu2026toposim}. This shift makes cost a central bottleneck, and serving work lowers per-call inference cost through quantization, speculative decoding, routing, and memory management \citep{frantar2023gptq,leviathan2023fast,chen2023frugalgpt,sheng2023flexgen,kwon2023efficient}. However, cost-reduction strategies often trade off LLM anchoring or error control: learned proxies move the transition rule into trained models, grouped representatives or static propagation leave unqueried agents dependent on local approximations, and serving-level optimizations reduce per-call cost without defining a recurrent population-transition estimator with propagation-bias diagnostics.

Compared with these lines of work, APS addresses the cost--fidelity gap by keeping the specified LLM as the online transition oracle while reducing online calls through adaptive prototype simulation. It queries core prototypes and singleton-tail agents, propagates same-round prototype responses through local response surfaces, and dynamically allocates prototype budgets using stratum size, residual error, and local similarity diagnostics. For error control, shadow-audit residual correction estimates propagation residuals for aggregate correction and future budget allocation, while tail-protected singleton routing directly queries selected isolated, heterogeneous, or high-curvature feature-space regions vulnerable to smoothing. Additional related-work details are discussed in Appendix~\ref{app:additional_related_work}.

\vspace{-0.1cm}

\section{Adaptive Prototype Simulation}
\label{sec:method}

\begin{figure}[t]
\centering
\includegraphics[width=\textwidth]{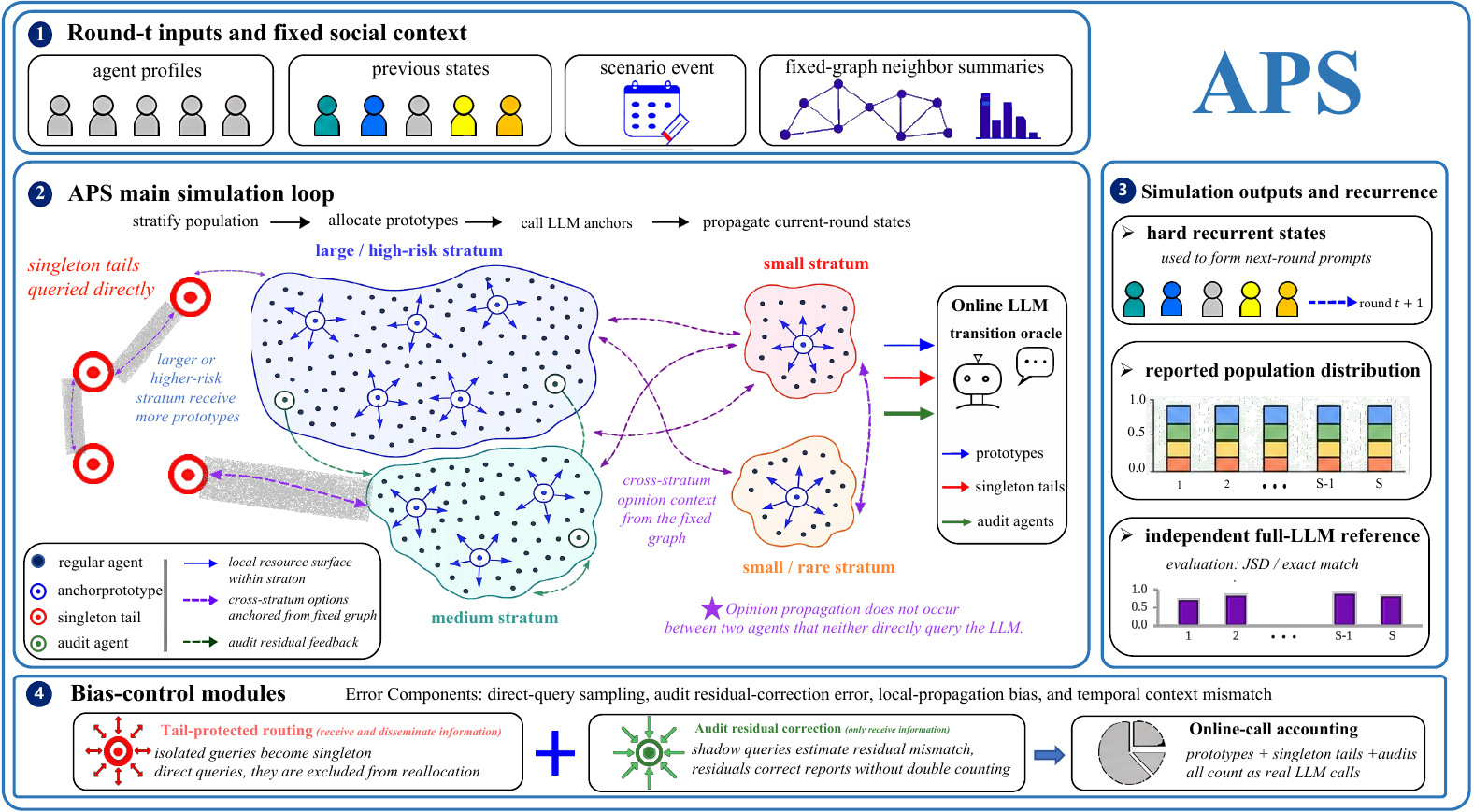}
\vspace{-0.4cm}
\caption{Overview of Adaptive Prototype Simulation (APS). Each round, APS builds prompts from profiles, scenario events, pre-round hard recurrent states, and fixed-graph summaries. It queries budgeted core-prototype, singleton-tail, and shadow-audit agents, then propagates prototype responses within core strata while fixed-graph summaries provide cross-stratum context. Hard recurrent states feed subsequent rounds, and the soft reporting ledger with shadow-audit residuals estimates the population distribution with online-call accounting.}
\vspace{-0.4cm}
\label{fig:framework}
\end{figure}

APS treats the specified LLM, prompt, scenario, option set, parser, and decoding rule as the online transition oracle, but queries only a selected subset of agents each round. The method section is organized as follows. Section~3.1 defines the problem setting, computational targets, and state records APS maintains; Section~3.2 gives an overview of the APS workflow, including prototype propagation and query allocation; Section~3.3 explains the two bias-control mechanisms, tail-protected singleton routing and shadow-audit residual correction; and Section~3.4 summarizes the resulting query complexity and error decomposition. Operational method details are collected in Appendix~\ref{app:aps_auxiliary_details}, and derivations are moved to Appendix~\ref{app:theory_details} so that the main text emphasizes the simulation logic.

\subsection{Problem Definition, Targets, and State Records}

We consider a population $\mathcal A_N=\{a_i\}_{i=1}^N$ of $N$ agents and a finite decision set $\mathcal Y$. Agent $i$ has static profile features $x_i$ and a position in a fixed social graph $G$. At round $t$, the scenario event is $s_t$, and the prompt for an agent includes its profile, its previous decision, and a summary of previous-round neighbor decisions. We write $K_\theta(y\mid z,s_t)$ for the decision distribution induced by the fixed LLM pipeline under prompt context $z$; here $\theta$ denotes the fixed model, prompt, parser, and decoding configuration. Deterministic decoding is the point-mass special case.

A full-LLM reference simulation, abbreviated as the full-reference simulation, queries every agent at every round. Its context for agent $i$ is $z_i^{\star,t}$, built from the agent profile, the full-reference previous state, and the corresponding neighbor summary, and its realized decision is $y_i^{\star,t}$. The expected full-reference population transition $\bar p_t^\star$ and the realized empirical reference distribution $p_t^\star$ are
\begin{equation}
\label{eq:full_reference_targets}
    \bar p_t^\star(y)=N^{-1}\sum_i K_\theta(y\mid z_i^{\star,t},s_t),
    \qquad
    p_t^\star(y)=N^{-1}\sum_i\mathbb I[y_i^{\star,t}=y].
\end{equation}

\vspace{-0.4cm}

APS cannot query every $z_i^{\star,t}$. Instead, it maintains an approximate hard recurrent state $\hat y_i^t$ for each agent and builds APS-context prompts $\hat z_i^t$ from these states. If every agent were queried under these APS-maintained contexts, the target would be $\bar p_t^A(y)=N^{-1}\sum_i K_\theta(y\mid \hat z_i^t,s_t)$. The gap between $\bar p_t^A$ and $\bar p_t^\star$ is the temporal context mismatch from approximate recurrence. In the API experiments, we observe realized categorical decisions, not calibrated option probabilities, so reference-aligned metrics compare against the empirical distribution $p_t^\star$. The full-reference simulation is a computational target for the specified LLM system, not a claim about true human behavior.

APS therefore keeps two role-specific records. The hard recurrent ledger ${\hat y_i^t}$ advances future prompts and neighbor summaries through APS-maintained contexts. The soft reporting ledger ${h_i^t}$ stores a per-agent, per-category probability vector over $\mathcal Y$: one-hot LLM decisions for directly queried agents and interpolated response-surface probabilities for propagated agents. The reported population distribution is the simplex-projected audit-corrected aggregate built from the soft ledger and design-weighted shadow-audit residuals, not the uncorrected empirical count of hard recurrent states.

\subsection{APS Workflow Overview}

Figure~\ref{fig:framework} shows the APS workflow. Before the first round, APS selects a singleton-tail set $O_N$ for tail-protected singleton routing and partitions the remaining agents into core strata $\{C_m\}$.\footnote{In the reported experiments, core strata are fixed because agents' profile features are treated as static. In settings where profile features evolve, APS should include an explicit profile-transition mechanism and recompute feature-space tail scores and core strata every round or after a fixed number of rounds.} At round $t$, it builds prompts from the pre-round hard recurrent states, chooses prototype sets $S_{m,t}\subset C_m$, queries the LLM for those prototypes and the singleton-tail agents, and propagates prototype decisions to non-prototype core agents through a local response surface. The online calls in a round thus consist of core-prototype queries, singleton-tail queries, and the shadow-audit queries defined below; only the first two update recurrent hard states. Detailed step-by-step pseudocode is given in Appendix~\ref{app:aps_algorithm_details}.

For a direct rollout agent, namely a core prototype or singleton-tail agent, the soft vector $h_i^t$ is the one-hot encoding of the parsed LLM decision. For a non-prototype core agent $j\in C_m\setminus S_{m,t}$, let $P_\kappa(j,t)\subseteq S_{m,t}$ be the set of up to $\kappa$ nearest queried prototypes within the same stratum, and let $w_{ji,t}$ be nonnegative interpolation weights with $\sum_{i\in P_\kappa(j,t)}w_{ji,t}=1$. APS sets
\begin{equation}
\label{eq:local_response_surface}
    h_j^t(y)=\sum_{i\in P_\kappa(j,t)} w_{ji,t}\,\mathbb I[\hat y_i^t=y],
    \qquad
    \hat y_j^t=\arg\max_{y\in\mathcal Y}h_j^t(y).
\end{equation}
Ties are broken by a fixed rule. This separation lets APS use hard states for recurrence while keeping soft local-response-surface information for aggregate reporting and shadow-audit residual correction.

The query budget is controlled by sublinear scale schedules. Core strata grow as $M_{\mathrm{core}}(N)=O(N^\eta)$ and the singleton-tail set grows as $M_{\mathrm{out}}(N)=O(N^\zeta)$, with sublinear exponents $\eta,\zeta<1$, so absolute counts grow while fractions shrink. The core prototype rate follows
\begin{equation}
\label{eq:sampling}
    \alpha(N)=
    \begin{cases}
    \alpha_b, & N\le N_b,\\
    \alpha_b(N_b/N)^\lambda, & N>N_b,
    \end{cases}
    \qquad 0<\lambda<1.
\end{equation}
Here $N_b$ is the baseline scale, $\alpha_b$ the baseline prototype rate, and $\lambda$ the decay exponent. For $N>N_b$, the queried fraction decreases but the absolute number of core prototypes grows.

If $B_t^{\mathrm{core}}=\lceil\alpha(N)N\rceil$ denotes the nominal pre-round core-prototype budget and $R_{m,t-1}$ is the previous residual-risk score defined in Section~\ref{sec:bias_controlled_propagation} with $R_{m,0}=1$, the continuous allocation for stratum $C_m$ is
\begin{equation}
\label{eq:budget_update}
    \widetilde B_{m,t}=
    B_t^{\mathrm{core}}
    \frac{|C_m|\sqrt{R_{m,t-1}+\tau}}
    {\sum_{\ell}|C_\ell|\sqrt{R_{\ell,t-1}+\tau}},
\end{equation}
where the denominator sums over all core strata to normalize allocation weights and $\tau>0$ stabilizes strata with near-zero scores. The rollout obtains the integer budget $B_{m,t}$ by rounding $\widetilde B_{m,t}$ and applying the minimum-one rule for nonempty strata described in Appendix~\ref{app:scale_schedules}.

\subsection{Bias-Controlled Routing and Auditing}
\label{sec:bias_controlled_propagation}

Prototype propagation is efficient only if local similarity is informative. APS therefore adds two safeguards that target distinct smoothing and residual-error failure modes.

\textbf{Tail-protected singleton routing.}
Local smoothing can distort isolated or heterogeneous feature regions when no close prototype is queried. APS therefore uses a feature-space tail score $s_i$ for each agent and directly queries the top-scoring $M_{\mathrm{out}}(N)$ agents every round. Here $s_i$ is a robust standardized distance from the feature median; Appendix~\ref{app:stratification_response_surface_tails} gives the exact scoring rule. These singleton-tail agents contribute one-hot soft vectors for themselves, but they are not used as prototype supports for other agents during local interpolation within core response surfaces. This reduces smoothing risk for selected isolated regions; it does not claim to identify all minority opinions.

\textbf{Shadow-audit residual correction.}
After propagation, APS audits a subset of non-prototype core agents. These shadow-audit queries use the same LLM, prompt format, scenario event, and APS-maintained context as rollout calls, but their labels are shadow observations: they do not enter the prototype set and do not overwrite current hard states. Let $U_t$ be the audit set, $\tilde y_i^t$ the shadow-audit label for audited agent $i$, and $\psi_{i,t}$ the known probability that this agent is included in the shadow-audit sample. APS reports

\vspace{-0.6cm}

\begin{equation}
\label{eq:aps_audit_corrected_estimator}
    \widehat p_t^{\mathrm{audit}}(y)
    =
    \frac{1}{N}\sum_{i=1}^{N} h_i^t(y)
    +
    \frac{1}{N}\sum_{i\in U_t}
    \frac{\mathbb I[\tilde y_i^t=y]-h_i^t(y)}{\psi_{i,t}}.
\end{equation}
The first term is the full-population local-response-surface baseline; the second is a design-weighted residual correction from shadow-audit queries. For JSD reporting, APS clips and renormalizes it to the probability simplex. With positive shadow-audit inclusion probabilities, the local response surface need not be correct for the unprojected estimator to be design-valid for the APS-context target; poor propagation instead appears as larger residual variance and later shadow-audit demand. The shadow-audit sampling rule and allocation diagnostics are given in Appendices~\ref{app:aps_algorithm_details} and~\ref{app:audit_diagnostics}.

Shadow-audit queries also update the next-round residual-risk score used in Eq.~\ref{eq:budget_update}. For stratum $C_m$, APS combines normalized shadow-audit diagnostics as

\vspace{-0.2cm}

\begin{equation}
\label{eq:residual_risk_score}
    R_{m,t}=\widehat V_{m,t}^{\mathrm{res}}
    +\lambda_\rho \widehat L_{m,t}^{2}\widehat\rho_{m,t}^{2}
    +\lambda_e \widehat e_{m,t}^{2}
    +\lambda_r(1-\widehat r_{m,t})^{2}.
\end{equation}

\vspace{-0.1cm}

Here $\widehat V_{m,t}^{\mathrm{res}}$ is residual variance, $\widehat\rho_{m,t}$ is the average distance from shadow-audited non-prototypes to their prototype supports, $\widehat L_{m,t}$ is a local disagreement or curvature estimate, $\widehat e_{m,t}$ is the shadow-audit mismatch rate, and $\widehat r_{m,t}$ is rare-state recall. Together, they form the residual-risk score used for later allocation. The weights $\lambda_\rho,\lambda_e,\lambda_r$ tune the relative contribution of these diagnostics.

\subsection{Complexity and Error Analysis}
\label{sec:targets_complexity_error}

The analysis concerns computational fidelity to the LLM-induced population transition, not its behavioral validity for real human populations. Counting core-prototype queries, singleton-tail queries, and shadow-audit queries, APS reduces the $O(TN)$ calls of full simulation to
\begin{equation}
\label{eq:aps_complexity}
C_{\mathrm{APS}}(N,T)=
O\!\left(T\left[N^{1-\lambda}+M_{\mathrm{core}}(N)+M_{\mathrm{out}}(N)+A(N)\right]\right)
\end{equation}
where $T$ is the number of rounds. Here $A(N)$ denotes the per-round order of the shadow-audit schedule $A_t(N)$, and the $M_{\mathrm{core}}(N)$ term accounts for the per-stratum minimum prototype-query floor. Under the production choice $A(N)=O(N^{1-\lambda})$, this becomes $O(T[N^{1-\lambda}+M_{\mathrm{tot}}(N)])$ with $M_{\mathrm{tot}}(N)=M_{\mathrm{core}}(N)+M_{\mathrm{out}}(N)$, sublinear in $N$ whenever the schedule exponents are below one.

The error decomposition tracks the four sources used across the paper: prototype-coverage error $\epsilon_{\mathrm{cov},t}$, finite shadow-audit residual-correction error $\epsilon_{\mathrm{audit},t}$, core local-propagation bias $\epsilon_{\mathrm{prop},t}$, and temporal context mismatch $\epsilon_{\mathrm{ctx},t}$. The first three form the current-round APS-context error; the fourth arises as hard recurrent states condition future prompts. If $e_t^\star$ is the induced full-reference error, then
\begin{equation}
\label{eq:error_decomposition}
    e_t^\star
    \le
    \epsilon_{\mathrm{cov},t}
    +\epsilon_{\mathrm{audit},t}
    +\epsilon_{\mathrm{prop},t}
    +\epsilon_{\mathrm{ctx},t}.
\end{equation}
Shadow-audit residual correction targets local-propagation bias in the unprojected aggregate estimator, rather than directly repairing the propagated hard states themselves, but poor propagation can still raise finite-audit variance and affect later rounds through hard recurrent states. Appendix~\ref{app:theory_details} gives the formal bounds, including design validity, residual-variance comparison, tail-bias removal, local-response-surface regularity, multi-round recurrence, and residual-aware allocation.

\vspace{-0.2cm}

\section{Experiments and Results}

\noindent This section first describes the experimental setup and then addresses the following research questions:
\begin{itemize}[labelindent=0pt, leftmargin=*, labelsep=0.35em,
                itemsep=0pt, topsep=0pt, parsep=0pt, partopsep=0pt]
    \item \textbf{RQ1:} How accurately and efficiently does APS match references at 10M-agent scale?
    \item \textbf{RQ2:} Under matched budgets, does APS outperform approximation baselines?
    \item \textbf{RQ3:} How much do residual correction, singleton routing, and adaptive allocation help?
    \item \textbf{RQ4:} Does APS avoid monotone error accumulation across simulation rounds?
    \item \textbf{RQ5:} Does APS remain robust across scenarios, LLM backends, and stratification backends?
\end{itemize}

\vspace{-0.1cm}

\subsection{Experimental Setup}

\textbf{Scenario and populations.}
We evaluate APS on an 8-round public-opinion simulation about a hypothetical subway chemical attack. Each round provides new information and asks agents to choose among five options; the task tests macro-level opinion dynamics rather than crisis policy. Agent profiles start from about 90K real respondents in the multi-country World Values Survey (WVS) \citep{wvs2022wave7}, represented with 19 standardized survey-derived features. These profiles are simulation inputs, not city-population claims. Larger populations use type-aware, stratum-preserving perturbations around real WVS records; Appendix~\ref{app:population_construction} gives the construction protocol.

\textbf{Methods and operating point.}
We evaluate APS against independent full-LLM reference rollouts and compare it with LS-Surrogate adapted from Light Society \citep{guan2025earthscale} and TopoSim-Coord adapted from TopoSim \citep{xu2026toposim}. Unless otherwise stated, APS uses $N_b=5000$, $\alpha_b=0.15$, baseline core-stratum count $M_b=10$, $\lambda=0.6$, $\eta=0.5$, $\zeta=0.4$, and $\beta_a=1-\lambda=0.4$; Appendix~\ref{app:table1_aps_settings} lists the Table~\ref{tab:validation} superscript settings. The 3K--100K rows use OpenRouter with glm-4.7-flash under a one-process four-key configuration. The 1M and 10M rows use the official Zhipu API with glm-4-flash; this provider split is an API-cost constraint, as the large-scale runs use the no-cost official Zhipu endpoint available to us. All GLM runs use deterministic decoding; Appendix~\ref{app:reproducibility} gives API settings, runtime and cost accounting, and Appendix~\ref{app:prompts} gives prompt text.

\textbf{Independent full-LLM reference rollouts.}
Full-LLM reference rollouts are independent of APS: they update the corresponding population through all rounds with their own recurrent states and neighbor summaries, and no APS state, propagated label, audit result, or budget decision enters the reference. These rollouts are computational targets, not evidence of human behavioral validity; protocol and accounting details are in Appendices~\ref{app:evaluation_protocol}--\ref{app:runtime_accounting}.

\textbf{Metrics.}
For reference-aligned evaluations, we report Jensen--Shannon divergence (JSD), exact match, and LLM calls. JSD compares the reported distribution with the empirical distribution of the corresponding full-LLM reference rollout; for APS, this uses the simplex-projected audit-corrected estimator $\widehat p_t^{\mathrm{audit}}$. Exact match measures aligned agreement between each method's recurrent hard top-1 state and the reference label, so it is auxiliary. Confidence intervals follow Appendix~\ref{app:confidence_intervals}.

\textbf{Result structure.}
The main text reports scale-oriented baselines (Table~\ref{tab:validation}), same-budget approximations (Table~\ref{tab:same_budget}), component and allocation ablations (Tables~\ref{tab:audit_outlier_ablation} and~\ref{tab:ablation}), a 10K full-LLM trajectory drift check (Figure~\ref{fig:full10k_drift}), and robustness tests across scenarios and backends (Figure~\ref{fig:generalization} and Table~\ref{tab:robustness_combined}).

\vspace{-0.1cm}

\subsection{Final-Round Accuracy, Cost, and 10M Scaling (RQ1)}

Table~\ref{tab:validation} is the main provider-aware scale comparison against LS-Surrogate \citep{guan2025earthscale} and TopoSim-Coord \citep{xu2026toposim}. Reference outputs are used for JSD and exact-match reporting; LS-Surrogate calls count reference-derived distillation labels, whereas APS uses no reference label for propagation or adaptation. APS calls include core-prototype, singleton-tail, and shadow-audit queries, and reductions divide brute-force full-simulation calls by reported method calls. Appendix~\ref{app:reproducibility} gives implementation, reference, call-accounting, runtime, and cost details.

\begin{table*}[t]
\centering
\caption{Final-round JSD, exact-match accuracy, and scaling. Superscripts on APS denote small-scale validation ($^\dagger$), high-budget validation ($^\ddagger$), and production schedule ($^\star$). APS calls include core-prototype, singleton-tail, and shadow-audit queries, but exclude independent reference construction; LS-Surrogate calls report counted reference-derived distillation labels; brackets report 95\% intervals.}
\label{tab:validation}
\fontsize{8.9pt}{9.5pt}\selectfont
\renewcommand{\arraystretch}{0.86}
\setlength{\aboverulesep}{0.25ex}
\setlength{\belowrulesep}{0.25ex}
\setlength{\tabcolsep}{4pt}
\renewcommand{\tabularxcolumn}[1]{m{#1}}
\begin{tabularx}{\textwidth}{llrrr>{\centering\arraybackslash}Xr>{\centering\arraybackslash}X}
\toprule
Scale & Method & LLM Calls & Reduction & JSD $\downarrow$ & JSD CI & Exact $\uparrow$ & Exact CI \\
\midrule
\rowcolor{APSTableGreen}
3K & APS$^{\dagger}$ & 8.9K & \textbf{2.7$\times$} & \textbf{0.011} & [0.0065, 0.0155] & \textbf{0.673} & [0.6560, 0.6896] \\
\validationdash
10K & LS-Surrogate & 8.0K & \textbf{10.0$\times$} & 0.166 & [0.1570, 0.1750] & 0.490 & [0.4802, 0.4998] \\
10K & TopoSim-Coord & 20.7K & 3.9$\times$ & 0.150 & [0.1414, 0.1586] & 0.344 & [0.3348, 0.3534] \\
\rowcolor{APSTableBlue}
10K & APS$^{\ddagger}$ & 20.8K & 3.8$\times$ & \textbf{0.014} & [0.0112, 0.0268] & \textbf{0.662} & [0.6527, 0.6712] \\
\validationdash
100K & LS-Surrogate & 80.0K & 10.0$\times$ & 0.179 & [0.1761, 0.1819] & 0.486 & [0.4829, 0.4891] \\
100K & TopoSim-Coord & 52.2K & \textbf{15.3$\times$} & 0.153 & [0.1503, 0.1557] & 0.338 & [0.3351, 0.3409] \\
\rowcolor{APSTableBlue}
100K & APS$^{\ddagger}$ & 172.0K & 4.7$\times$ & \textbf{0.021} & [0.0199, 0.0221] & \textbf{0.639} & [0.6360, 0.6420] \\
\validationdash
1M & LS-Surrogate & 800.0K & 10.0$\times$ & 0.223 & [0.2220, 0.2240] & 0.450 & [0.4490, 0.4510] \\
1M & TopoSim-Coord & 130.2K & 61.4$\times$ & 0.172 & [0.1711, 0.1729] & 0.334 & [0.3331, 0.3349] \\
\rowcolor{APSTableLavender}
1M & APS$^{\star}$ & 83.4K & \textbf{95.9$\times$} & \textbf{0.062} & [0.0614, 0.0626] & \textbf{0.517} & [0.5160, 0.5180] \\
\validationdash
10M & LS-Surrogate & 8.0M & 10.0$\times$ & 0.264 & [0.2637, 0.2643] & 0.411 & [0.4107, 0.4113] \\
10M & TopoSim-Coord & 329.7K & 242.6$\times$ & 0.191 & [0.1907, 0.1913] & 0.325 & [0.3247, 0.3253] \\
\rowcolor{APSTableLavender}
10M & APS$^{\star}$ & 209.9K & \textbf{381.1$\times$} & \textbf{0.094} & [0.0938, 0.0942] & \textbf{0.544} & [0.5437, 0.5443] \\
\bottomrule

\vspace{-0.5cm}

\end{tabularx}
\end{table*}

The accuracy trend is consistent across matched scales: where baselines are reported, APS has the lowest final-round JSD and the highest exact-match diagnostic. At 10K agents, for example, APS reduces JSD from 0.166 for LS-Surrogate and 0.150 for TopoSim-Coord to 0.014. At 10M agents under the production schedule, APS obtains JSD 0.094 with 95\% CI [0.0938, 0.0942], compared with 0.264 for LS-Surrogate and 0.191 for TopoSim-Coord.

\begin{figure*}[t]
  \centering
  \begin{minipage}[t]{0.49\textwidth}
  \vspace{0pt}
  \centering
  \includegraphics[width=\linewidth]{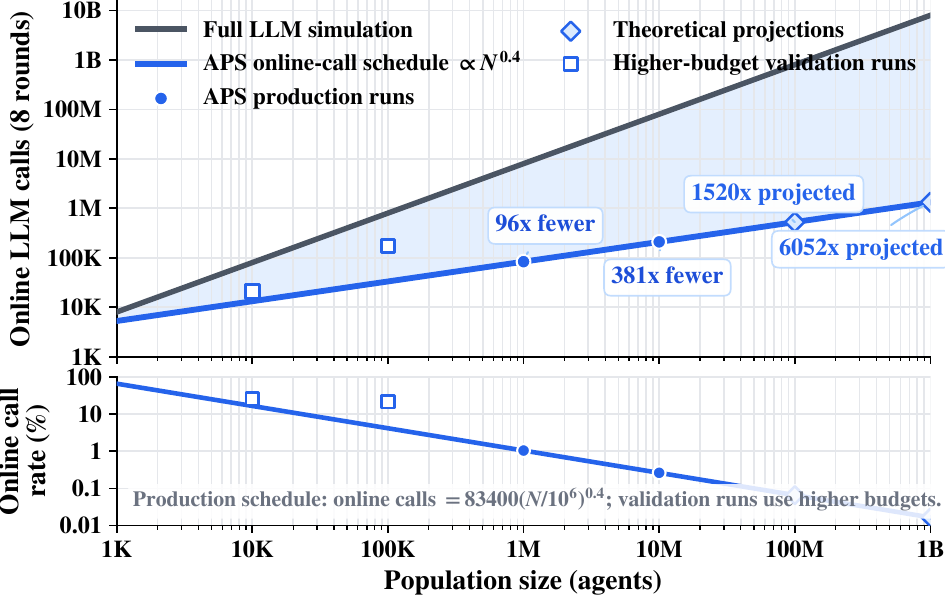}
  \par\vspace{-5pt}
  \refstepcounter{figure}\label{fig:cost_savings}
  {\fontsize{8.0pt}{8.7pt}\selectfont\noindent\parbox[t]{\linewidth}{\normalfont\noindent Figure~\thefigure. LLM-call scaling for brute-force full simulation and APS over 1K--1B agents. Filled circles mark completed production runs up to 10M, diamonds show theoretical 100M and 1B schedule projections, and squares mark
smaller-scale high-budget validation runs.\par}}
  \end{minipage}\hfill
  \begin{minipage}[t]{0.49\textwidth}
  \vspace{0pt}
  \centering
  \fontsize{7.35pt}{8.15pt}\selectfont
  \setlength{\tabcolsep}{4.0pt}
  \renewcommand{\arraystretch}{1.32}
  \begin{tabular}{@{}lccc@{}}
  \toprule
  Method & Calls & JSD & Exact \\
  \midrule
  \mbox{Strat. sample \citep{cochran1977sampling}} & 15.5K & 0.088 & 0.515 \\
  \mbox{Cluster assign. \citep{lloyd1982least}} & 15.5K & 0.167 & 0.321 \\
  \mbox{Label prop. \citep{zhu2003semi}} & 15.5K & 0.061 & 0.513 \\
  \mbox{Core-set \citep{feldman2020coresets}} & 15.5K & 0.124 & 0.408 \\
  \mbox{Strat. emp. \citep{cochran1977sampling}} & 15.5K & 0.016 & 0.539 \\
  \mbox{Active sel. \citep{xia2024less}} & 15.5K & 0.033 & 0.514 \\
  \mbox{Medoid \citep{kaufman1990finding}} & 15.5K & 0.029 & 0.605 \\
  \mbox{Online surr. \citep{diaw2024surrogates}} & 15.5K & 0.017 & 0.589 \\
  \rowcolor{APSTableBlue}[0pt][0pt]
  \mbox{APS} & 15.5K & \textbf{0.011} & \textbf{0.624} \\
  \bottomrule
  \end{tabular}
  \begingroup
  \renewcommand{\APScaptionfont}{\fontsize{8.0pt}{8.7pt}\selectfont}
  \sidecaptionof{table}{Same-budget 10K approximation-baseline comparison. Calls include all online LLM calls used by each method. Lower JSD is better; exact match uses recurrent hard top-1 states aligned to the full-LLM reference rollout.}
  \label{tab:same_budget}
  \endgroup
  \end{minipage}
  \vspace{-0.6cm}
\end{figure*}

Figure~\ref{fig:cost_savings} separates the cost trend. At 10M agents over eight rounds, brute-force full simulation requires 80M LLM calls, whereas APS uses 209.9K counted online calls, including core, singleton-tail, and audit queries, a $381.1\times$ reduction over brute force; under the same Zhipu configuration, wall-clock time is 26.64 hours for APS\footnote{For APS, wall-clock time is dominated by prototype clustering and simulation-pipeline processing, including feature processing, propagation, audit aggregation, checkpointing, and bookkeeping; API waiting is a smaller share under this multi-key setup.} and 9.56 days for the full-LLM reference rollout. These scale-oriented budgets are not same-budget baselines, and LS-Surrogate's low inference cost follows counted task-specific distillation rather than online oracle querying during rollout.

\vspace{-0.1cm}

\subsection{Same-Budget Approximation Baselines (RQ2)}

Table~\ref{tab:same_budget} fixes the 10K scale and 15.5K online-call budget over eight rounds, then compares APS with sampling, propagation, selection, coreset, and surrogate approximations under the same online-call budget, population, scenario, parser, LLM endpoint, and reference task, isolating allocation and propagation choices rather than scale, prompt, or reference effects. APS has the lowest final JSD, the primary distributional metric, and highest exact match, an auxiliary hard-state diagnostic, in this controlled comparison; Appendix~\ref{app:same_budget_baselines} gives implementation details.

\vspace{-0.1cm}

\subsection{Component and Adaptive Allocation Ablations (RQ3)}

Table~\ref{tab:audit_outlier_ablation} reports 10K component checks for the two bias-control modules. At matched 12.3K calls, the two modules reduce JSD from 0.075 to 0.019 and 0.022 to 0.011, respectively, while raising exact match from 0.584 to 0.659 and 0.591 to 0.662. The lower-call component-off settings show the same direction, with larger errors when either module is disabled; Appendix~\ref{app:ablation_protocols} gives protocol details.

\begin{figure*}[t]
\centering
\begin{minipage}[t]{0.520\textwidth}
\vspace{0pt}
\centering
\fontsize{7.3pt}{8.0pt}\selectfont
\setlength{\tabcolsep}{1.0pt}
\renewcommand{\arraystretch}{1.10}
\begin{tabularx}{\linewidth}{@{}*{6}{>{\centering\arraybackslash}X}@{}}
\toprule
Rate & Allocation & Seeds & Calls & JSD & Exact \\
\midrule
0.05 & fixed & 2 & 5.3K & 0.064 & 0.373 \\
\rowcolor{APSTableBlue}
0.05 & adaptive & 2 & 5.3K & 0.021 & 0.489 \\
0.10 & fixed & 2 & 6.5K & 0.056 & 0.392 \\
\rowcolor{APSTableBlue}
0.10 & adaptive & 2 & 6.5K & 0.018 & 0.506 \\
0.15 & fixed & 2 & 7.7K & 0.052 & 0.424 \\
\rowcolor{APSTableBlue}
0.15 & adaptive & 2 & 7.7K & 0.020 & 0.542 \\
0.20 & fixed & 2 & 8.9K & 0.047 & 0.457 \\
\rowcolor{APSTableBlue}
0.20 & adaptive & 2 & 8.9K & 0.018 & 0.625 \\
0.30 & fixed & 2 & 11.3K & 0.042 & 0.506 \\
\rowcolor{APSTableBlue}
0.30 & adaptive & 2 & 11.3K & 0.011 & 0.638 \\
0.50 & fixed & 2 & 16.2K & 0.026 & 0.507 \\
\rowcolor{APSTableBlue}
0.50 & adaptive & 2 & 16.2K & 0.006 & 0.664 \\
\bottomrule
\end{tabularx}
\sidecaptionof{table}{Adaptive allocation ablation at 3K scale, averaged over two seeds. Calls include prototype, singleton-tail, and audit components.}
\label{tab:ablation}
\end{minipage}
\hfill
\begin{minipage}[t]{0.455\textwidth}
\vspace{0pt}
\centering
\includegraphics[width=\linewidth,trim=0 1.4pt 0 0,clip]{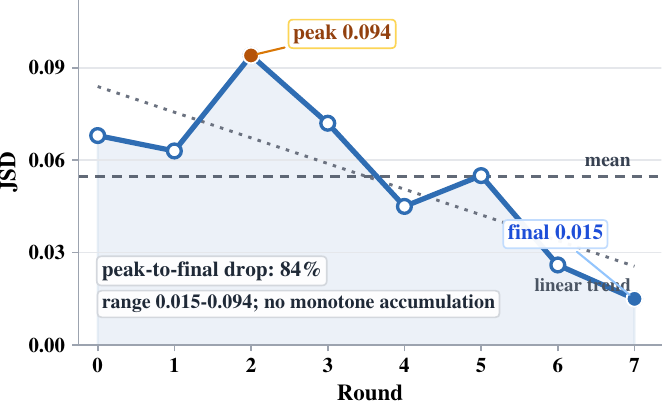}
\sidecaptionof{figure}{Per-round JSD in the completed 10K full-LLM trajectory check, with mean and linear-trend diagnostics; the nonmonotone path is consistent with no monotone accumulation in this check.}
\label{fig:full10k_drift}
\end{minipage}

\vspace{-0.4cm}

\end{figure*}

Table~\ref{tab:ablation} compares adaptive and fixed allocation at matched 3K budgets, counting prototype, singleton-tail, and audit calls. This isolates allocation from total-query effects: adaptive allocation lowers JSD at every rate and reaches exact match 0.664; fixed allocation remains at or below 0.507.

\begin{table}[t]
  \centering
  \caption{10K ablations for shadow-audit residual correction and tail-protected singleton routing.}
  \label{tab:audit_outlier_ablation}
  \begin{adjustbox}{max width=\textwidth}
  \begin{tabular}{llclll}
  \toprule
  Variant & LLM Calls & JSD $\downarrow$ & JSD 95\% CI & Exact match $\uparrow$ & Exact match 95\% CI \\
  \midrule
  \rowcolor{gray!12}[0pt][0pt]
  \multicolumn{6}{l}{\textit{Shadow-audit residual correction}} \\
  Base APS without residual correction & 9.3K & 0.098 & [0.090, 0.106] & 0.536 & [0.526, 0.546] \\
  APS without residual correction (same budget) & 12.3K & 0.075 & [0.066, 0.082] & 0.584 & [0.573, 0.591] \\
  APS with residual correction & 12.3K & 0.019 & [0.014, 0.024] & 0.659 & [0.650, 0.668] \\
  \midrule
  \rowcolor{gray!12}[0pt][0pt]
  \multicolumn{6}{l}{\textit{Tail-protected singleton routing}} \\
  APS without singleton-tail routing & 9.4K & 0.030 & [0.022, 0.038] & 0.529 & [0.519, 0.539] \\
  APS without singleton-tail routing (same budget) & 12.3K & 0.022 & [0.017, 0.028] & 0.591 & [0.584, 0.599] \\
  APS with singleton-tail routing & 12.3K & 0.011 & [0.007, 0.019] & 0.662 & [0.654, 0.671] \\
  \bottomrule
  \end{tabular}
  \end{adjustbox}

  \vspace{-0.4cm}

\end{table}

\vspace{-0.2cm}

\subsection{Multi-Round Drift  (RQ4)}

\vspace{-0.1cm}

An independent 10K full-LLM trajectory reference queries all matched agents for eight rounds under the same scenario stages. Figure~\ref{fig:full10k_drift} tests whether APS introduces severe cross-round error accumulation: the JSD path is nonmonotone, peaks at 0.094, and ends at 0.015, providing no evidence of monotone round-by-round accumulation in this check.

\vspace{-0.2cm}

\subsection{Robustness Across Scenarios, LLM Backends, and Stratification Backends (RQ5)}

\vspace{-0.1cm}

\label{sec:robustness}

\begin{wrapfigure}{r}{0.60\textwidth}
\vspace{-1.1em}
\centering
\includegraphics[width=\linewidth]{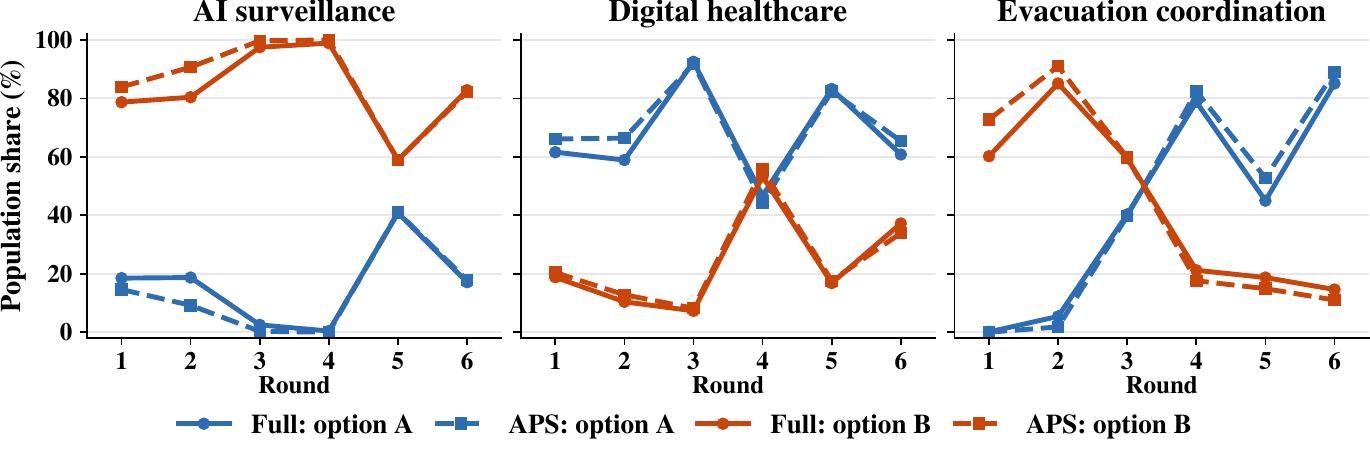}\par
\vspace*{0cm}
{\setlength{\abovecaptionskip}{0pt}
\caption{Multi-scenario trajectory visualization. Each panel tracks two scenario-specific focal options for full-LLM and APS rollouts; final distributional checks appear in Table~\ref{tab:robustness_combined}.}
\label{fig:generalization}}
\vspace{-1.0em}
\end{wrapfigure}
Table~\ref{tab:robustness_combined} reports compact 10K robustness checks across scenario prompts, LLM backends, and stratification backends, tested on matched agents against paired full-LLM references, and Figure~\ref{fig:generalization} plots representative scenario trajectories. These checks do not establish universal robustness, but they suggest that the observed cost--fidelity tradeoff is not confined to the main crisis-opinion prompt or a single LLM or stratification backend within the tested design space; Appendix~\ref{app:robustness_results} gives protocol notes.

\begin{table}[t]
\centering
\caption{Scenario, LLM backend, and stratification backend robustness. Scenario rows use glm-4.7-flash; LLM backend rows report means over the three 10K scenario checks; stratification rows use the same 10K full-LLM reference. Metrics are reference-aligned within each check. Brackets are 95\% intervals.}
\vspace{-0.1cm}
\label{tab:robustness_combined}
\label{tab:scenario_generalization}
\label{tab:model_generalization}
\label{tab:clustering_backend}
\fontsize{7.8pt}{8.25pt}\selectfont
\setlength{\tabcolsep}{3.8pt}
\renewcommand{\arraystretch}{0.88}
\begin{tabular*}{\textwidth}{@{\extracolsep{\fill}}l>{\raggedright\arraybackslash}p{0.30\textwidth}cccc@{}}
\toprule
Check & Variant & JSD $\downarrow$ & JSD 95\% CI & Exact match $\uparrow$ & Exact 95\% CI \\
\midrule
Scenario & Polarized AI surveillance & 0.004 & [0.001, 0.016] & 0.820 & [0.812, 0.827] \\
Scenario & Minority-sensitive digital healthcare & 0.027 & [0.008, 0.067] & 0.632 & [0.623, 0.641] \\
Scenario & Coordination evacuation & 0.025 & [0.008, 0.053] & 0.850 & [0.843, 0.857] \\
\midrule
LLM backend & deepseek/deepseek-v4-flash & 0.007 & [0.005, 0.011] & 0.806 & [0.801, 0.810] \\
LLM backend & google/gemini-3-flash-preview & 0.008 & [0.006, 0.011] & 0.832 & [0.828, 0.836] \\
LLM backend & openai/gpt-5.4-nano & 0.004 & [0.002, 0.007] & 0.753 & [0.748, 0.758] \\
LLM backend & qwen/qwen3.6-flash & 0.002 & [0.001, 0.004] & 0.805 & [0.800, 0.809] \\
LLM backend & x-ai/grok-4.1-fast & 0.007 & [0.005, 0.010] & 0.849 & [0.845, 0.853] \\
LLM backend & xiaomi/mimo-v2-omni & 0.011 & [0.008, 0.015] & 0.656 & [0.651, 0.661] \\
\midrule
Stratification & Gaussian mixture, fixed $K$ (14 strata) & 0.055 & [0.044, 0.070] & 0.609 & [0.599, 0.619] \\
Stratification & BIRCH, fixed $K$ (14 strata) & 0.050 & [0.039, 0.064] & 0.592 & [0.582, 0.602] \\
Stratification & BIRCH, threshold only (32 strata) & 0.039 & [0.029, 0.050] & 0.594 & [0.584, 0.604] \\
\bottomrule
\end{tabular*}
\vspace{-0.1cm}
\end{table}

\vspace{-0.1cm}

\section{Limitations}
APS evaluates computational fidelity to an LLM-induced transition target, not behavioral validity as a model of human behavior. Biased or miscalibrated reference LLM outputs may therefore be preserved at scale: shadow-audit residual correction and tail-protected singleton routing reduce approximation bias relative to the online transition oracle, but do not validate that oracle. Although independent full-LLM references avoid leakage from APS states, external validity remains limited by WVS-derived seed records, perturbed profiles, generated social graphs, and fixed scenario prompts. In particular, the city-scale scenarios in this paper should be read as urban stress tests rather than calibrated models of any specific city: the current population, graph, and prompts do not encode city-specific census composition, neighborhood geography, mobility or commuting flows, local media exposure, or institutional histories. Broader domains, longer horizons, richer and geographically grounded networks, and empirical human or city-level observational data remain important future tests; Appendix~\ref{app:ethics} discusses dual-use risks, subgroup reporting, and governance requirements.

\vspace{-0.1cm}

\section{Discussion and Conclusion}

\textbf{Discussion.}
APS frames population-scale LLM-agent simulation as recurrent LLM-oracle allocation, not only serving-level acceleration. For fixed online-call budgets, APS allocates core-prototype, singleton-tail, and shadow-audit queries, keeps the specified LLM as the online transition oracle, and uses queried prototype responses as same-round semantic anchors for local response surfaces. This clarifies the fidelity target: APS approximates the LLM-induced population transition against independent full-LLM references, reports the soft reporting ledger's audit-corrected aggregate $\widehat p_t^{\mathrm{audit}}$, and maintains hard recurrent states for later prompts. Exact match remains an auxiliary hard-state diagnostic; shadow-audit residual correction neither repairs every propagated state nor eliminates all temporal context mismatch. These bias-control mechanisms are complementary: tail-protected singleton routing directly queries selected isolated, heterogeneous, or high-curvature feature-space agents; shadow-audit residual correction estimates propagation residuals for aggregate correction and residual-aware allocation. Scalable LLM social simulators should specify the computational transition target, make key assumptions explicit, count all online LLM calls, separate recurrent states from reported estimands, expose residual and tail-coverage diagnostics, and retain independent full-LLM references plus positive-probability shadow-audit calls where feasible.

\textbf{Conclusion.}
We presented APS, a bias-controlled framework for population-scale LLM-agent simulation. APS keeps the LLM as the online transition oracle, queries adaptive core prototypes and singleton-tail agents, and uses shadow-audit residuals to correct population estimates and guide future budget allocation. Counting core-prototype, singleton-tail, and shadow-audit calls, APS reduces full-simulation cost from $O(NT)$ to $O(T(N^{1-\lambda}+M_{\mathrm{tot}}(N)))$. Empirically, APS scales to a 10M-agent, 8-round public-opinion simulation with 209.9K online LLM calls, a $381.1\times$ reduction over full simulation, and final-round JSD 0.094 against the corresponding full-LLM reference. Beyond scaling, APS separates the computational transition target from behavioral validity, making cost--fidelity tradeoffs auditable. Same-budget comparisons, component ablations, the completed 10K trajectory check, and compact robustness checks support the claim: APS can expand LLM-agent simulation scale while maintaining low distributional discrepancy to the full-LLM computational target.

\bibliographystyle{plainnat}
\bibliography{references}

@book{epstein1996growing,
  title={Growing artificial societies: social science from the bottom up},
  author={Epstein, Joshua M and Axtell, Robert},
  year={1996}
}

@article{macy2002factors,
  title={From factors to actors: Computational sociology and agent-based modeling},
  author={Macy, Michael W and Willer, Robert},
  journal={Annual review of sociology},
  volume={28},
  number={1},
  pages={143--166},
  year={2002}
}

@article{castellano2009statistical,
  title={Statistical physics of social dynamics},
  author={Castellano, Claudio and Fortunato, Santo and Loreto, Vittorio},
  journal={Reviews of modern physics},
  volume={81},
  number={2},
  pages={591--646},
  year={2009}
}

@article{kirman1992representative,
  title={Whom or what does the representative individual represent?},
  author={Kirman, Alan P},
  journal={Journal of economic perspectives},
  volume={6},
  number={2},
  pages={117--136},
  year={1992}
}

@book{cochran1977sampling,
  title={Sampling techniques},
  author={Cochran, William Gemmell},
  year={1977}
}

@article{horvitz1952generalization,
  title={A generalization of sampling without replacement from a finite universe},
  author={Horvitz, Daniel G and Thompson, Donovan J},
  journal={Journal of the American statistical Association},
  volume={47},
  number={260},
  pages={663--685},
  year={1952}
}

@article{lloyd1982least,
  title={Least squares quantization in PCM},
  author={Lloyd, Stuart},
  journal={IEEE transactions on information theory},
  volume={28},
  number={2},
  pages={129--137},
  year={1982}
}

@inproceedings{park2023generative,
  title={Generative agents: Interactive simulacra of human behavior},
  author={Park, Joon Sung and O'Brien, Joseph and Cai, Carrie Jun and Morris, Meredith Ringel and Liang, Percy and Bernstein, Michael S},
  booktitle={Proceedings of the 36th annual acm symposium on user interface software and technology},
  pages={1--22},
  year={2023}
}

@article{wang2024surveyagents,
  title={A survey on large language model based autonomous agents},
  author={Wang, Lei and Ma, Chen and Feng, Xueyang and Zhang, Zeyu and Yang, Hao and Zhang, Jingsen and Chen, Zhiyuan and Tang, Jiakai and Chen, Xu and Lin, Yankai and others},
  journal={Frontiers of Computer Science},
  volume={18},
  number={6},
  pages={186345},
  year={2024}
}

@inproceedings{guo2024multiagent,
  title={Large language model based multi-agents: A survey of progress and challenges},
  author={Guo, Taicheng and Chen, Xiuying and Wang, Yaqi and Chang, Ruidi and Pei, Shichao and Chawla, Nitesh V and Wiest, Olaf and Zhang, Xiangliang},
  journal={arXiv preprint arXiv:2402.01680},
  year={2024}
}

@article{xi2023rise,
  title={The rise and potential of large language model based agents: A survey},
  author={Xi, Zhiheng and Chen, Wenxiang and Guo, Xin and He, Wei and Ding, Yiwen and Hong, Boyang and Zhang, Ming and Wang, Junzhe and Jin, Senjie and Zhou, Enyu and others},
  journal={Science China Information Sciences},
  volume={68},
  number={2},
  pages={121101},
  year={2025}
}

@inproceedings{li2023camel,
  title={Camel: Communicative agents for" mind" exploration of large language model society},
  author={Li, Guohao and Hammoud, Hasan and Itani, Hani and Khizbullin, Dmitrii and Ghanem, Bernard},
  journal={Advances in neural information processing systems},
  volume={36},
  pages={51991--52008},
  year={2023}
}

@inproceedings{chen2024agentverse,
  title={Agentverse: Facilitating multi-agent collaboration and exploring emergent behaviors},
  author={Chen, Weize and Su, Yusheng and Zuo, Jingwei and Yang, Cheng and Yuan, Chenfei and Chan, Chi-Min and Yu, Heyang and Lu, Yaxi and Hung, Yi-Hsin and Qian, Chen and others},
  booktitle={The Twelfth International Conference on Learning Representations},
  year={2023}
}

@inproceedings{hong2024metagpt,
  title={MetaGPT: Meta programming for a multi-agent collaborative framework},
  author={Hong, Sirui and Zhuge, Mingchen and Chen, Jonathan and Zheng, Xiawu and Cheng, Yuheng and Wang, Jinlin and Zhang, Ceyao and Wang, Zili and Yau, Steven Ka Shing and Lin, Zijuan and others},
  booktitle={The twelfth international conference on learning representations},
  year={2023}
}

@article{argyle2023out,
  title={Out of one, many: Using language models to simulate human samples},
  author={Argyle, Lisa P and Busby, Ethan C and Fulda, Nancy and Gubler, Joshua R and Rytting, Christopher and Wingate, David},
  journal={Political Analysis},
  volume={31},
  number={3},
  pages={337--351},
  year={2023}
}

@article{horton2023large,
  title={Large language models as simulated economic agents: What can we learn from homo silicus?},
  author={Horton, John J and Filippas, Apostolos and Manning, Benjamin S},
  year={2023},
  institution={National Bureau of Economic Research}
}

@inproceedings{aher2023using,
  title={Using large language models to simulate multiple humans and replicate human subject studies},
  author={Aher, Gati V and Arriaga, Rosa I and Kalai, Adam Tauman},
  booktitle={International conference on machine learning},
  pages={337--371},
  year={2023},
  organization={PMLR}
}

@article{ashery2025emergent,
  title={Emergent social conventions and collective bias in LLM populations},
  author={Ashery, Ariel Flint and Aiello, Luca Maria and Baronchelli, Andrea},
  journal={Science Advances},
  volume={11},
  number={20},
  pages={eadu9368},
  year={2025}
}

@article{grossmann2023transformation,
  title={AI and the transformation of social science research},
  author={Grossmann, Igor and Feinberg, Matthew and Parker, Dawn C and Christakis, Nicholas A and Tetlock, Philip E and Cunningham, William A},
  journal={Science},
  volume={380},
  number={6650},
  pages={1108--1109},
  year={2023}
}

@article{bail2024socialscience,
  title={Can generative AI improve social science?},
  author={Bail, Christopher A},
  journal={Proceedings of the National Academy of Sciences},
  volume={121},
  number={21},
  pages={e2314021121},
  year={2024}
}

@article{davidson2025integrating,
  title={Integrating generative artificial intelligence into social science research: Measurement, prompting, and simulation},
  author={Davidson, Thomas and Karell, Daniel},
  journal={Sociological Methods \& Research},
  volume={54},
  number={3},
  pages={775--793},
  year={2025}
}

@article{ziems2024transform,
  title={Can large language models transform computational social science?},
  author={Ziems, Caleb and Held, William and Shaikh, Omar and Chen, Jiaao and Zhang, Zhehao and Yang, Diyi},
  journal={Computational Linguistics},
  volume={50},
  number={1},
  pages={237--291},
  year={2024}
}

@article{dillion2023replace,
  title={Can AI language models replace human participants?},
  author={Dillion, Danica and Tandon, Niket and Gu, Yuling and Gray, Kurt},
  journal={Trends in Cognitive Sciences},
  volume={27},
  number={7},
  pages={597--600},
  year={2023}
}

@article{chen2023economic,
  title={The emergence of economic rationality of GPT},
  author={Chen, Yiting and Liu, Tracy Xiao and Shan, You and Zhong, Songfa},
  journal={Proceedings of the National Academy of Sciences},
  volume={120},
  number={51},
  pages={e2316205120},
  year={2023}
}

@article{qu2024publicopinion,
  title={Performance and biases of large language models in public opinion simulation},
  author={Qu, Yao and Wang, Jue},
  journal={Humanities and Social Sciences Communications},
  volume={11},
  number={1},
  pages={1--13},
  year={2024}
}

@article{bisbee2024synthetic,
  title={Synthetic replacements for human survey data? The perils of large language models},
  author={Bisbee, James and Clinton, Joshua D and Dorff, Cassy and Kenkel, Brenton and Larson, Jennifer M},
  journal={Political Analysis},
  volume={32},
  number={4},
  pages={401--416},
  year={2024}
}

@article{abdurahman2025primer,
  title={A primer for evaluating large language models in social-science research},
  author={Abdurahman, Suhaib and Salkhordeh Ziabari, Alireza and Moore, Alexander K and Bartels, Daniel M and Dehghani, Morteza},
  journal={Advances in Methods and Practices in Psychological Science},
  volume={8},
  number={2},
  pages={25152459251325174},
  year={2025}
}

@article{wang2025replace,
  title={Large language models that replace human participants can harmfully misportray and flatten identity groups},
  author={Wang, Angelina and Morgenstern, Jamie and Dickerson, John P},
  journal={Nature Machine Intelligence},
  volume={7},
  number={3},
  pages={400--411},
  year={2025}
}

@article{lin2025six,
  title={Six fallacies in substituting large language models for human participants},
  author={Lin, Zhicheng},
  journal={Advances in Methods and Practices in Psychological Science},
  volume={8},
  number={3},
  pages={25152459251357566},
  year={2025}
}

@article{manning2024automated,
  title={Automated social science: Language models as scientist and subjects},
  author={Manning, Benjamin S and Zhu, Kehang and Horton, John J},
  year={2024},
  institution={National Bureau of Economic Research}
}

@article{yang2024oasis,
  title={Oasis: Open agent social interaction simulations with one million agents},
  author={Yang, Ziyi and Zhang, Zaibin and Zheng, Zirui and Jiang, Yuxian and Gan, Ziyue and Wang, Zhiyu and Ling, Zijian and Chen, Jinsong and Ma, Martz and Dong, Bowen and others},
  journal={arXiv preprint arXiv:2411.11581},
  year={2024}
}

@article{gao2024s3,
  title={S3: Social-network simulation system with large language model-empowered agents},
  author={Gao, Chen and Lan, Xiaochong and Lu, Zhihong and Mao, Jinzhu and Piao, Jinghua and Wang, Huandong and Jin, Depeng and Li, Yong},
  journal={arXiv preprint arXiv:2307.14984},
  year={2023}
}

@article{park2024generative1000,
  title={Generative agent simulations of 1,000 people},
  author={Park, Joon Sung and Zou, Carolyn Q and Shaw, Aaron and Hill, Benjamin Mako and Cai, Carrie and Morris, Meredith Ringel and Willer, Robb and Liang, Percy and Bernstein, Michael S},
  journal={arXiv preprint arXiv:2411.10109},
  volume={52},
  year={2024}
}

@article{mou2024individual,
  title={From individual to society: A survey on social simulation driven by large language model-based agents},
  author={Mou, Xinyi and Ding, Xuanwen and He, Qi and Wang, Liang and Liang, Jingcong and Zhang, Xinnong and Sun, Libo and Lin, Jiayu and Zhou, Jie and Xuanjing, Huang and others},
  journal={ACM Computing Surveys},
  year={2024}
}

@inproceedings{zhou2024sotopia,
  title={Sotopia: Interactive evaluation for social intelligence in language agents},
  author={Zhou, Xuhui and Zhu, Hao and Mathur, Leena and Zhang, Ruohong and Yu, Haofei and Qi, Zhengyang and Morency, Louis-Philippe and Bisk, Yonatan and Fried, Daniel and Neubig, Graham and others},
  journal={arXiv preprint arXiv:2310.11667},
  year={2023}
}

@inproceedings{frantar2023gptq,
  title={Optq: Accurate quantization for generative pre-trained transformers. 2023},
  author={Frantar, E and Ashkboos, S and Hoefler, T and Alistarh, D},
  year={2023}
}

@inproceedings{leviathan2023fast,
  title={Fast inference from transformers via speculative decoding},
  author={Leviathan, Yaniv and Kalman, Matan and Matias, Yossi},
  booktitle={International Conference on Machine Learning},
  pages={19274--19286},
  year={2023},
  organization={PMLR}
}

@article{chen2023frugalgpt,
  title={Frugalgpt: How to use large language models while reducing cost and improving performance},
  author={Chen, Lingjiao and Zaharia, Matei and Zou, James},
  journal={arXiv preprint arXiv:2305.05176},
  year={2023}
}

@inproceedings{sheng2023flexgen,
  title={Flexgen: High-throughput generative inference of large language models with a single gpu},
  author={Sheng, Ying and Zheng, Lianmin and Yuan, Binhang and Li, Zhuohan and Ryabinin, Max and Chen, Beidi and Liang, Percy and R{\'e}, Christopher and Stoica, Ion and Zhang, Ce},
  booktitle={International Conference on Machine Learning},
  pages={31094--31116},
  year={2023},
  organization={PMLR}
}

@inproceedings{kwon2023efficient,
  title={Efficient memory management for large language model serving with pagedattention},
  author={Kwon, Woosuk and Li, Zhuohan and Zhuang, Siyuan and Sheng, Ying and Zheng, Lianmin and Yu, Cody Hao and Gonzalez, Joseph and Zhang, Hao and Stoica, Ion},
  booktitle={Proceedings of the 29th symposium on operating systems principles},
  pages={611--626},
  year={2023}
}

@article{guan2025earthscale,
  title={Modeling earth-scale human-like societies with one billion agents},
  author={Guan, Haoxiang and He, Jiyan and Fan, Liyang and Ren, Zhenzhen and He, Shaobin and Yu, Xin and Chen, Yuan and Zheng, Shuxin and Liu, Tie-Yan and Liu, Zhen},
  journal={arXiv preprint arXiv:2506.12078},
  year={2025}
}

@article{chopra2025largepopulationmodels,
  title={Large population models},
  author={Chopra, Ayush},
  journal={arXiv preprint arXiv:2507.09901},
  year={2025}
}

@inproceedings{tang2024gensim,
  title={Gensim: A general social simulation platform with large language model based agents},
  author={Tang, Jiakai and Gao, Heyang and Pan, Xuchen and Wang, Lei and Tan, Haoran and Gao, Dawei and Chen, Yushuo and Chen, Xu and Lin, Yankai and Li, Yaliang and others},
  booktitle={Proceedings of the 2025 Conference of the Nations of the Americas Chapter of the Association for Computational Linguistics: Human Language Technologies (System Demonstrations)},
  pages={143--150},
  year={2025}
}

@article{piao2025agentsociety,
  title={Agentsociety: Large-scale simulation of llm-driven generative agents advances understanding of human behaviors and society},
  author={Piao, Jinghua and Yan, Yuwei and Zhang, Jun and Li, Nian and Yan, Junbo and Lan, Xiaochong and Lu, Zhihong and Zheng, Zhiheng and Wang, Jing Yi and Zhou, Di and others},
  year={2025}
}

@article{xu2026toposim,
  title={Topology-Aware LLM-Driven Social Simulation: A Unified Framework for Efficient and Realistic Agent Dynamics},
  author={Xu, Yuwei and Zhang, Shulun and Zhou, Yingli and Zeng, Shipei and Lakshmanan, Laks VS and Ma, Chenhao},
  journal={arXiv preprint arXiv:2604.18011},
  year={2026}
}

@article{vezhnevets2023concordia,
  title={Generative agent-based modeling with actions grounded in physical, social, or digital space using Concordia},
  author={Vezhnevets, Alexander Sasha and Agapiou, John P and Aharon, Avia and Ziv, Ron and Matyas, Jayd and Du{\'e}{\~n}ez-Guzm{\'a}n, Edgar A and Cunningham, William A and Osindero, Simon and Karmon, Danny and Leibo, Joel Z},
  journal={arXiv preprint arXiv:2312.03664},
  year={2023}
}

@article{feldman2020coresets,
  title={Core-sets: Updated survey},
  author={Feldman, Dan},
  booktitle={Sampling techniques for supervised or unsupervised tasks},
  pages={23--44},
  year={2019}
}

@book{kaufman1990finding,
  title={Finding groups in data: an introduction to cluster analysis},
  author={Kaufman, Leonard and Rousseeuw, Peter J},
  year={2009}
}

@inproceedings{zhu2003semi,
  title={Semi-supervised learning using gaussian fields and harmonic functions},
  author={Zhu, Xiaojin and Ghahramani, Zoubin and Lafferty, John D},
  booktitle={Proceedings of the 20th International conference on Machine learning (ICML-03)},
  pages={912--919},
  year={2003}
}

@book{settles2009active,
  title={Active learning literature survey},
  author={Settles, Burr},
  year={2009}
}

@book{sarndal1992model,
  title={Model assisted survey sampling},
  author={S{\"a}rndal, Carl-Erik and Swensson, Bengt and Wretman, Jan},
  year={2003}
}

@article{gao2024llmabm,
  title={Large language models empowered agent-based modeling and simulation: A survey and perspectives},
  author={Gao, Chen and Lan, Xiaochong and Li, Nian and Yuan, Yuan and Ding, Jingtao and Zhou, Zhilun and Xu, Fengli and Li, Yong},
  journal={Humanities and Social Sciences Communications},
  volume={11},
  number={1},
  pages={1--24},
  year={2024}
}

@article{li2023laennet,
  title={LaenNet: Learning robust GCNs by propagating labels},
  author={Zhang, Chunxu and Li, Ximing and Pei, Hongbin and Zhang, Zijian and Liu, Bing and Yang, Bo},
  journal={Neural Networks},
  volume={168},
  pages={652--664},
  year={2023}
}

@inproceedings{diao2024activeprompt,
  title={Active prompting with chain-of-thought for large language models},
  author={Diao, Shizhe and Wang, Pengcheng and Lin, Yong and Pan, Rui and Liu, Xiang and Zhang, Tong},
  booktitle={Proceedings of the 62nd Annual Meeting of the Association for Computational Linguistics (Volume 1: Long Papers)},
  pages={1330--1350},
  year={2024}
}

@inproceedings{xia2024less,
  title={Less: Selecting influential data for targeted instruction tuning},
  author={Xia, Mengzhou and Malladi, Sadhika and Gururangan, Suchin and Arora, Sanjeev and Chen, Danqi},
  journal={arXiv preprint arXiv:2402.04333},
  year={2024}
}

@article{mukherjee2023stratified,
  title={Stratified sampling: some associated problems},
  author={Mukherjee, SP},
  journal={Calcutta Statistical Association Bulletin},
  volume={75},
  number={1},
  pages={48--59},
  year={2023}
}

@article{wesolowski2024recursive,
  title={Recursive Neyman algorithm for optimum sample allocation under box constraints on sample sizes in strata},
  author={Wieczorkowski, Robert and W{\u{A}}{\l}jciak, Wojciech and others},
  journal={arXiv preprint arXiv:2304.07034},
  year={2023}
}

@article{diaw2024surrogates,
  title={Efficient learning of accurate surrogates for simulations of complex systems},
  author={Diaw, Abdourahmane and McKerns, Michael and Sagert, Irina and Stanton, Liam G and Murillo, Michael S},
  journal={Nature Machine Intelligence},
  volume={6},
  number={5},
  pages={568--577},
  year={2024}
}

@misc{wvs2022wave7,
  title={World values survey wave 7 (2017-2022) cross-national data-set},
  author={Haerpfer, Christian and Inglehart, Ronald and Moreno, Alejandro and Welzel, Christian and Kizilova, Kseniya and Diez-Medrano, Jaime and Lagos, Marta and Norris, Pippa and Ponarin, Eduard and Puranen, Bi},
  journal={(No Title)},
  year={2022}
}

@article{predhumeau2023synthetic,
  title={A synthetic population for agent-based modelling in Canada},
  author={Pr{\'e}dhumeau, Manon and Manley, Ed},
  journal={Scientific Data},
  volume={10},
  number={1},
  pages={148},
  year={2023}
}

@article{jiang2024synthetic,
  title={A large-scale geographically explicit synthetic population with social networks for the united states},
  author={Jiang, Na and Yin, Fuzhen and Wang, Boyu and Crooks, Andrew T},
  journal={Scientific Data},
  volume={11},
  number={1},
  pages={1204},
  year={2024}
}

\clearpage
\appendix

\AppendixContentsPage

\noindent\textbf{Appendix roadmap.}

\noindent\textbf{Appendix roadmap.}
Appendix~\ref{app:additional_related_work} expands the connections to LLM social simulators, interdisciplinary simulation uses, prototype propagation, model-assisted estimation, and adaptive allocation. Appendix~\ref{app:aps_auxiliary_details} contains method-level details: detailed APS algorithms, shadow-audit diagnostics, scale schedules, and Table~\ref{tab:validation} APS settings. Appendix~\ref{app:theory_details} gives the derivations behind query complexity, audit correction, error decomposition, and residual-aware allocation. Appendix~\ref{app:reproducibility} collects experimental protocol details: baseline implementations, reference construction, call accounting, runtime records, confidence intervals, ablation protocols, population and graph construction, LLM-call logging, shadow-audit inclusion probabilities, and scale resources. Appendix~\ref{app:prompts} gives the exact prompt templates and scenario text. Appendix~\ref{app:robustness_results} records robustness-check protocol notes for Section~\ref{sec:robustness}. Appendix~\ref{app:ethics} gives use boundaries and social-impact considerations.

\section{Additional Related Work}
\label{app:additional_related_work}

\textbf{LLM agents as a simulation substrate.}
Recent surveys and perspective articles describe a broader shift from using LLMs as task solvers to using them as components in synthetic social worlds, behavioral experiments, and computational social-science workflows \citep{xi2023rise,wang2024surveyagents,guo2024multiagent,grossmann2023transformation,bail2024socialscience,davidson2025integrating,ziems2024transform,gao2024llmabm,mou2024individual}. Early generative-agent and multi-agent systems focus on believable memory, role play, interaction, social intelligence, and group collaboration \citep{park2023generative,li2023camel,chen2024agentverse,hong2024metagpt,zhou2024sotopia,vezhnevets2023concordia}. This line establishes that LLM agents can produce coherent local behavior, but it also shifts the methodological question from whether agents can act plausibly in small scenes to whether population-level trajectories remain valid, calibrated, and affordable when many agents interact over time.

\textbf{Interdisciplinary behavioral and social simulation.}
LLM-based simulation now intersects psychology, political science, economics, sociology, and computational social science. Studies use LLM personas or agent populations to replicate human-subject experiments, survey-like responses, public opinion, economic choice, trust, and convention formation \citep{argyle2023out,aher2023using,dillion2023replace,horton2023large,chen2023economic,qu2024publicopinion,park2024generative1000,ashery2025emergent}. At the same time, social-science and psychology work warns that synthetic respondents should not be treated as human truth: LLM outputs can be biased, overconfident, prompt-sensitive, temporally stale, or weak at preserving subgroup heterogeneity and causal structure \citep{bisbee2024synthetic,abdurahman2025primer,wang2025replace,lin2025six}. APS follows the more conservative view that LLM simulations are model-based instruments requiring explicit reference definitions, validation metrics, and error controls rather than replacements for human evidence.

\textbf{Population-scale platforms and decision-support use cases.}
The practical motivation for large LLM-agent simulation is increasingly framed around scenario screening, public-opinion analysis, platform governance, policy prototyping, crisis response, and other ``what-if'' uses where researchers or decision makers need macro-level trajectories before acting in the real world \citep{manning2024automated,qu2024publicopinion,piao2025agentsociety}. Scale-oriented systems simulate social networks, online platforms, synthetic populations, and large societies using general simulation engines, social-media environments, learned or optimized surrogates, differentiable population frameworks, or topology-aware grouping \citep{gao2024s3,tang2024gensim,yang2024oasis,predhumeau2023synthetic,jiang2024synthetic,guan2025earthscale,chopra2025largepopulationmodels,xu2026toposim}. These systems broaden the application space, but they also make cost--fidelity accounting central: a result meant to inform real decisions must report what was queried, what was propagated, what reference it approximates, and where approximation bias may enter.

\textbf{Scaling strategies and their statistical risks.}
Scaling LLM-agent simulation can reduce cost through model-serving optimizations, smaller or learned surrogates, grouped representatives, static or topology-aware propagation, and platform-level batching \citep{frantar2023gptq,leviathan2023fast,chen2023frugalgpt,sheng2023flexgen,kwon2023efficient,guan2025earthscale,xu2026toposim}. These strategies address different parts of the cost problem, but they can also introduce distinct statistical risks: removing the online LLM anchor changes the transition rule, grouping agents can smooth isolated or heterogeneous regions, and recurrent propagation can accumulate context mismatch across rounds. APS is closest to the scale-oriented simulation line, but differs by preserving the specified LLM as the online transition oracle and pairing prototype propagation with online residual correction, tail-protected singleton routing, and explicit call accounting.

\textbf{Prototype propagation and bias control.}
Representative-agent models, coresets, and graph-based label propagation show how sparse labels or representative points can summarize larger populations \citep{kirman1992representative,feldman2020coresets,zhu2003semi,li2023laennet}. These approaches are effective when local structure is informative, but their usual objective is static approximation or label completion rather than recurrent LLM-agent simulation. APS draws on this prototype view in a setting where propagated hard recurrent states condition later prompts, so prototype propagation must be paired with tail-protected singleton routing, shadow-audit residual correction, and a soft reporting ledger distinct from the hard recurrent ledger.

\textbf{Auditing, model-assisted estimation, and adaptive allocation.}
Model-assisted survey estimation uses a prediction model to form an initial population estimate and probability samples to estimate the residual \citep{horvitz1952generalization,cochran1977sampling,sarndal1992model}. Classical stratified and Neyman allocation study how to distribute samples across strata, with recent work extending allocation under practical constraints \citep{mukherjee2023stratified,wesolowski2024recursive}. LLM work on active prompting and influential data selection similarly motivates selective querying under limited oracle budgets \citep{settles2009active,diao2024activeprompt,xia2024less}. In APS, shadow-audit residual correction occurs online inside a multi-round LLM-agent simulation: shadow labels correct the reported aggregate estimator and guide future budget allocation, but do not overwrite hard recurrent states.

\section{APS Algorithm Details, Diagnostics, and Scale Schedules}
\label{app:aps_auxiliary_details}

This appendix collects the operational method details omitted from the main method for space: detailed pseudocode, shadow-audit diagnostics, scale schedules, and the concrete APS settings used in Table~\ref{tab:validation}. These details support reproducibility and the derivations in Appendix~\ref{app:theory_details}, while the main text retains only the formulas needed to understand APS.

\subsection{Detailed APS Algorithms}
\label{app:aps_algorithm_details}
Algorithm~\ref{alg:aps} expands the main rollout loop. Algorithm~\ref{alg:aps_audit} expands the shadow-audit residual correction used after prototype propagation. We keep them separate because the first algorithm defines the recurrent simulation path, whereas the second defines the estimator correction and next-round allocation diagnostics.

\begin{algorithm}[h]
\small
\caption{Detailed APS Main Simulation Loop}
\label{alg:aps}
\begin{algorithmic}[1]
\STATE \textbf{Input:} agent features $\{x_i\}_{i=1}^N$, graph $G$, initial states $y^0$, scenario events $\{s_t\}_{t=1}^T$, option set $\mathcal Y$, LLM pipeline and parser, schedules $M_{\mathrm{core}}(N)$, $M_{\mathrm{out}}(N)$, $\alpha(N)$, shadow-audit schedule $A_t(N)$, support size $\kappa$, interpolation rule $\mathcal I$, random seed $\xi$.
\STATE \textbf{Output:} hard recurrent states $\{\hat y^t\}_{t=1}^T$, projected shadow-audit-corrected distributions, diagnostics, rollout-query records, and shadow-audit records.
\STATE Set seed $\xi$; initialize $\hat y_i^0=y_i^0$ and stratum scores $R_{m,0}=1$.
\STATE Compute tail scores $s_i$; select singleton-tail set $O_N$ of size $M_{\mathrm{out}}(N)$.
\STATE Partition $\mathcal A_N\setminus O_N$ into $M_{\mathrm{core}}(N)$ core strata $\{C_m\}$.
\FOR{$t=1$ to $T$}
    \STATE Build neighbor summaries $\hat g_i^t=\Gamma_i(G,\hat y^{t-1})$ for all agents.
    \STATE Compute the round core-prototype budget from $\alpha(N)N$ and allocate stratum budgets $\{B_{m,t}\}$ using $|C_m|$, previous scores $R_{m,t-1}$, and the minimum-one rounding rule.
    \STATE Select prototypes $S_{m,t}\subset C_m$ with $|S_{m,t}|=B_{m,t}$; set $S_t=\cup_m S_{m,t}$ and direct rollout set $D_t=S_t\cup O_N$.
    \STATE Query the LLM for every $i\in D_t$ using profile $x_i$, previous state $\hat y_i^{t-1}$, neighbor summary $\hat g_i^t$, and event $s_t$; parse the response into an option in $\mathcal Y$.
    \STATE Set $h_i^t$ to the one-hot parsed decision and $\hat y_i^t$ to that decision for every directly queried agent $i\in D_t$.
    \FOR{each core stratum $C_m$}
        \FOR{each non-prototype agent $j\in C_m\setminus S_{m,t}$}
            \STATE Find the $\kappa$ nearest queried prototypes in $S_{m,t}$ under the implementation distance.
            \STATE Form $h_j^t$ by interpolation rule $\mathcal I$ over their one-hot decisions.
            \STATE Set $\hat y_j^t=\arg\max_{y\in\mathcal Y} h_j^t(y)$ with deterministic tie-breaking.
        \ENDFOR
    \ENDFOR
    \STATE Run Algorithm~\ref{alg:aps_audit} on the non-prototype core correction frame.
    \STATE Store hard states, projected audit-corrected distribution, diagnostics, and call records.
\ENDFOR
\end{algorithmic}
\end{algorithm}

\begin{algorithm}[h]
\small
\caption{Detailed Shadow-Audit Correction and Score Update}
\label{alg:aps_audit}
\begin{algorithmic}[1]
\STATE \textbf{Input:} soft vectors $h^t$, previous hard states $\hat y^{t-1}$, graph summaries $\hat g^t$, correction frame $F_t=\cup_m(C_m\setminus S_{m,t})$, shadow-audit budget $A_t(N)$, shadow-audit rule, and diagnostic weights.
\STATE \textbf{Output:} unprojected and projected shadow-audit-corrected distributions, shadow-audit diagnostics, inclusion probabilities, and next-round scores $R_{m,t}$.
\STATE Allocate shadow-audit counts $a_{m,t}$ across core strata, respecting $0\le a_{m,t}\le |F_t\cap C_m|$ and $\sum_m a_{m,t}\le A_t(N)$.
\FOR{each core stratum $C_m$}
    \STATE Select audit agents $U_{m,t}\subseteq F_t\cap C_m$ using the recorded audit design, with inclusion probabilities $\psi_{i,t}>0$.
    \STATE Query each $i\in U_{m,t}$ with the same LLM, parser, event $s_t$, and APS context $(x_i,\hat y_i^{t-1},\hat g_i^t)$ used by rollout calls.
    \STATE Parse the shadow decision $\tilde y_i^t$ and record retry or repair metadata.
    \STATE Compute residual vectors $\mathbb I[\tilde y_i^t=y]-h_i^t(y)$ for all $y\in\mathcal Y$.
    \STATE Summarize mismatch, residual variance, monitoring JSD, prototype-support distance, local disagreement or curvature, and rare-state recall.
    \STATE Update $R_{m,t}$ from the normalized diagnostics using Eq.~\ref{eq:residual_risk_score}.
\ENDFOR
\STATE Form the unprojected audit-corrected estimator by Eq.~\ref{eq:aps_audit_corrected_estimator}.
\STATE Clip negative entries and renormalize to the simplex for JSD reporting; keep the unprojected vector for the design-validity analysis.
\STATE Return both estimates, diagnostics, inclusion probabilities, and updated scores.
\end{algorithmic}
\end{algorithm}

\subsection{Shadow-Audit Diagnostic Definitions}
\label{app:audit_diagnostics}
For stratum $C_m$ at round $t$, let $U_{m,t}\subset C_m\setminus S_t$ be a shadow-audit set sampled across predicted state groups, with extra coverage for rare states, and let $a_{m,t}=|U_{m,t}|$. Let $\hat y_i^t$ be the APS-predicted state and $\tilde y_i^t$ be the shadow-audit state for $i\in U_{m,t}$. The mismatch, residual-variance, and monitoring-JSD summaries are
\[
\begin{aligned}
    \hat e_{m,t}
    &=\frac{1}{a_{m,t}}\sum_{i\in U_{m,t}}\mathbb{I}[\hat y_i^t\neq \tilde y_i^t],\\
    \widehat V_{m,t}^{\mathrm{res}}
    &=\sum_{y\in\mathcal{Y}}\widehat{\mathrm{Var}}_{i\in U_{m,t}}\!\left(\mathbb{I}[\tilde y_i^t=y]-h_i^t(y)\right),\\
    \widehat{\mathrm{JSD}}_{m,t}
    &=\mathrm{JSD}\big(\hat p_{U_{m,t}},\tilde p_{U_{m,t}}\big).
\end{aligned}
\]
Here $h_i^t(y)$ is the local-response-surface probability, and $\widehat{\mathrm{JSD}}_{m,t}$ is retained as a shadow-audit monitoring statistic. The residual-risk score in the main text additionally uses $\hat\rho_{m,t}$, the average distance from shadow-audited non-prototypes to their queried-prototype support; $\widehat L_{m,t}$, a local disagreement-slope or curvature estimate; and $\hat r_{m,t}$, the fraction of shadow-audit rare-state labels in $U_{m,t}$ that are also recovered by APS.

\subsection{Scale Schedules}
\label{app:scale_schedules}
The default core-stratum schedule is
\begin{equation}
\label{eq:cluster_schedule}
    M_{\mathrm{core}}(N)=
    \begin{cases}
    M_b, & N\le N_b,\\
    \left\lfloor M_b\left(\frac{N}{N_b}\right)^{\eta}\right\rfloor, & N>N_b,
    \end{cases}
    \qquad N_b=5000,\quad M_b=10,\quad 0<\eta<1.
\end{equation}
The concrete tail schedule used in our experiments is
\begin{equation}
\label{eq:outlier_schedule}
    M_{\mathrm{out}}(N)=
    \begin{cases}
    \lceil \delta_0 N_b\rceil, & N\le N_b,\\
    \left\lceil \delta_0 N_b\left(\frac{N}{N_b}\right)^{\zeta}\right\rceil, & N>N_b,
    \end{cases}
    \qquad \delta_0=0.05,\quad 0<\zeta<1.
\end{equation}
The number of singleton-tail queries grows as $O(N^\zeta)$, while the routed fraction shrinks as $O(N^{\zeta-1})$. This fraction is an algorithmic coverage budget, not an estimate of feature-space tail prevalence.

The shadow-audit-call schedule is
\[
    A_t(N)=\max\left\{A_{\min},
    \left\lfloor \gamma N_b\max\left(1,\left(\frac{N}{N_b}\right)^{\beta_a}\right)\right\rfloor
    \right\},
    \qquad 0<\beta_a<1,
\]
where $\beta_a$ controls shadow-audit-call growth and $\gamma$ sets the shadow-audit-call coefficient. Experiments set $\gamma=0.05$, $\beta_a=1-\lambda$, and $A_{\min}=1$.

For the residual-aware allocation in Eq.~\ref{eq:budget_update}, the minimum-one rule assigns at least one core prototype to each nonempty core stratum whenever the nominal core-prototype budget permits; the remaining slots are distributed according to the residual-aware stratum weights and then rounded to integers.

\subsection{Scaling-Table APS Settings}
\label{app:table1_aps_settings}

Table~\ref{tab:aps_table1_settings} gives the concrete APS settings behind the three superscripts in Table~\ref{tab:validation}. All three use $T=8$, seed 42, robust-MAD tail scoring, mini-batch or streaming-compatible $K$-means for core strata, inverse-distance interpolation with $\kappa=5$, initial scores $R_{m,0}=1$, and the residual-aware allocation rule in Eq.~\ref{eq:residual_risk_score} with normalized diagnostics, $\lambda_\rho=\lambda_e=\lambda_r=1$, and $\tau=10^{-6}$.

\begin{table}[h]
\centering
\caption{Concrete APS schedule settings for Table~\ref{tab:validation}. Core rate determines the pre-round core-prototype budget before residual-aware allocation across core strata. Reported calls are completed online calls across eight rounds and include core-prototype, singleton-tail, and shadow-audit queries.}
\label{tab:aps_table1_settings}
\small
\begin{adjustbox}{max width=\textwidth}
\begin{tabular}{lllllll}
\toprule
Sup. & Table rows & Core rate & $M_{\mathrm{core}}$ & $M_{\mathrm{out}}$ & Shadow-audit $A_t$ & Calls \\
\midrule
$\dagger$ & 3K & fixed $0.20$ & $10$ & $250$ & $250$ & 8.9K \\
$\ddagger$ & 10K, 100K & fixed $0.20$ & $\lfloor10(N/5000)^{0.5}\rfloor$ & $\lceil250(N/5000)^{0.4}\rceil$ & $\max\{1,\lfloor250(N/5000)^{0.4}\rfloor\}$ & 20.8K; 172.0K \\
$\star$ & 1M, 10M & $0.15(5000/N)^{0.6}$ & $\lfloor10(N/5000)^{0.5}\rfloor$ & $\lceil250(N/5000)^{0.4}\rceil$ & $\max\{1,\lfloor250(N/5000)^{0.4}\rfloor\}$ & 83.4K; 209.9K \\
\bottomrule
\end{tabular}
\end{adjustbox}
\end{table}

\begin{figure}[h]
\centering
\includegraphics[width=0.9\textwidth]{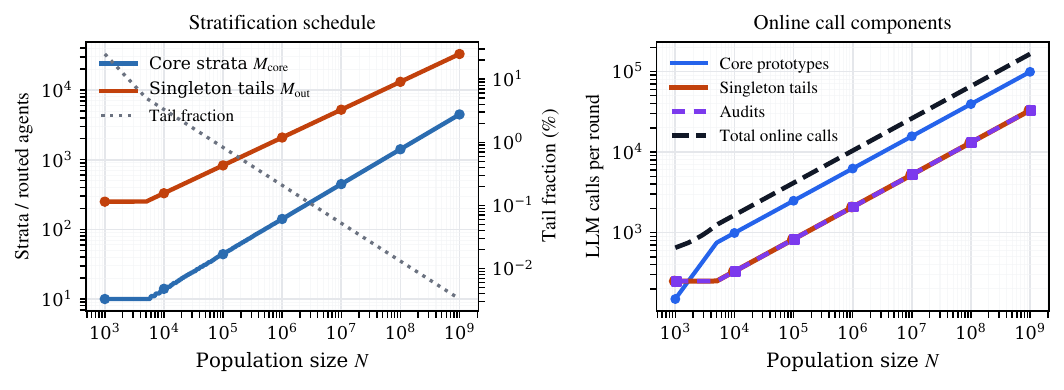}
\caption{Scale-dependent APS schedules. Left: core strata grow as $N^\eta$, singleton-tail agents grow as $N^\zeta$, and the routed fraction shrinks as $N^{\zeta-1}$. Right: per-round online LLM calls under the production schedule, counting core-prototype, singleton-tail, and shadow-audit queries.}
\label{fig:cluster_outlier_schedule}
\end{figure}

\section{Theory Details and Error Decomposition}
\label{app:theory_details}

This appendix gives the derivations summarized in Section~\ref{sec:method}.

\subsection{Query Complexity}
For $N>N_b$, Eq.~\ref{eq:sampling} gives the raw prototype budget
\[
    \alpha(N)N=\alpha_bN_b^\lambda N^{1-\lambda}=O(N^{1-\lambda}).
\]
The minimum-one floor in Algorithm~\ref{alg:aps} can add at most one core-prototype query per nonempty core stratum, giving $O(M_{\mathrm{core}}(N))$ calls. Tail-protected singleton routing queries $M_{\mathrm{out}}(N)$ agents per round, and APS adds $A(N)$ shadow-audit queries per round. Summing over $T$ rounds yields
\[
    C_{\mathrm{APS}}(N,T)
    =O\!\left(T[N^{1-\lambda}+M_{\mathrm{core}}(N)+M_{\mathrm{out}}(N)+A(N)]\right).
\]
Substituting $M_{\mathrm{core}}(N)=O(N^\eta)$, $M_{\mathrm{out}}(N)=O(N^\zeta)$, and $A(N)=O(N^{\beta_a})$ gives $O(TN^{\max\{1-\lambda,\eta,\zeta,\beta_a\}})$. With $\beta_a=1-\lambda$, the audit term has the same order as the prototype term, giving $O(T[N^{1-\lambda}+M_{\mathrm{tot}}(N)])$.

\subsection{Design Validity of Audit Correction}
Fix round $t$ and condition on $\mathcal F_{t-1}^{\mathrm{APS}}$. Suppose directly queried units satisfy $\mathbb E[\boldsymbol Z_i^t\mid\mathcal F_{t-1}^{\mathrm{APS}}]=\boldsymbol q_{i,t}^{A}$ for $i\in D_t$, audited correction-frame units satisfy
\[
    \mathbb E[\tilde{\boldsymbol Z}_i^t\mid\mathcal F_{t-1}^{\mathrm{APS}},D_t,\{\boldsymbol h_j^t\}_{j\in F_t}]
    =\boldsymbol q_{i,t}^{A},
\]
and the audit design has known positive inclusion probabilities with
\[
    \mathbb E\!\left[
    \frac{\mathbb I[i\in U_t]}{\psi_{i,t}}
    \mid\mathcal F_{t-1}^{\mathrm{APS}},D_t,\{\boldsymbol h_j^t\}_{j\in F_t}
    \right]=1.
\]
Directly queried agents contribute $\boldsymbol q_{i,t}^{A}$ in expectation. For $i\in F_t$, conditioning on the constructed response surface gives
\[
    \mathbb E\!\left[
    \frac{\mathbb I[i\in U_t]}{\psi_{i,t}}
    (\tilde{\boldsymbol Z}_i^t-\boldsymbol h_i^t)
    \right]
    =\boldsymbol q_{i,t}^{A}-\boldsymbol h_i^t.
\]
Adding the baseline $\boldsymbol h_i^t$ yields $\boldsymbol q_{i,t}^{A}$. Summing over direct and correction-frame units gives
\[
    \mathbb E[\widehat{\boldsymbol p}_t^{\mathrm{audit}}\mid\mathcal F_{t-1}^{\mathrm{APS}}]
    =\bar{\boldsymbol p}_t^{A}.
\]
The response surface need not be correct for unbiasedness; it affects variance. The simplex projection used for JSD is a finite-sample reporting transformation and can introduce projection bias, so the unbiasedness result applies before clipping and renormalization. A sufficient way to maintain positivity under adaptive shadow auditing is a mixture design
\[
    \psi_{i,t}=(1-\epsilon)\psi_{i,t}^{\mathrm{adaptive}}+\epsilon\psi_{i,t}^{\mathrm{explore}},
    \qquad \epsilon>0,
\]
where $\psi_{i,t}^{\mathrm{explore}}$ assigns nonzero probability to all units in the correction frame.

\subsection{Connection to the Full-Reference Target}
Let
\[
    \epsilon_{\mathrm{ctx},t}
    =\|\bar{\boldsymbol p}_t^{A}-\bar{\boldsymbol p}_t^{\star}\|_1
    =\left\|\frac{1}{N}\sum_{i=1}^{N}(\boldsymbol q_{i,t}^{A}-\boldsymbol q_{i,t}^{\star})\right\|_1.
\]
By design unbiasedness for $\bar{\boldsymbol p}_t^A$,
\[
    \left\|
    \mathbb E[\widehat{\boldsymbol p}_t^{\mathrm{audit}}\mid\mathcal F_{t-1}^{\mathrm{APS}}]
    -\bar{\boldsymbol p}_t^{\star}
    \right\|_1
    \le \epsilon_{\mathrm{ctx},t}.
\]
If the LLM-induced transition kernel is locally Lipschitz in a context metric $d_z$,
\[
    \|K_\theta(\cdot\mid z,s_t)-K_\theta(\cdot\mid z',s_t)\|_1\le L_t d_z(z,z'),
\]
then
\[
    \epsilon_{\mathrm{ctx},t}
    \le
    \frac{L_t}{N}\sum_{i=1}^{N}d_z(\hat z_i^t,z_i^{\star,t}),
    \qquad
    z_i^{\star,t}=(x_i,y_i^{\star,t-1},g_i^{\star,t}).
\]
Thus the gap to the full-reference target is governed by mismatch between APS-maintained histories and full-LLM histories, not by shadow-audit residual correction alone.

\subsection{Residual Variance Reduction}
Fix category $y$ and write $Z_i(y)=\mathbb I[\tilde y_i^t=y]$, $h_i(y)=h_i^t(y)$, and $q_i(y)=q_{i,t}^{A}(y)$. Under independent Poisson shadow-audit sampling, conditional on realized shadow-audit labels and the response surface,
\[
    \mathrm{Var}_{U_t}(\widehat p_t^{\mathrm{audit}}(y)\mid \tilde y,h)
    =\frac{1}{N^2}\sum_{i\in F_t}\frac{1-\psi_{i,t}}{\psi_{i,t}}(Z_i(y)-h_i(y))^2.
\]
The Horvitz--Thompson estimator corresponds to $h_i(y)=0$ and has design variance
\[
    \mathrm{Var}_{U_t}(\widehat p_t^{\mathrm{HT}}(y)\mid \tilde y)
    =\frac{1}{N^2}\sum_{i\in F_t}\frac{1-\psi_{i,t}}{\psi_{i,t}}Z_i(y)^2.
\]
APS therefore replaces raw-label variation by residual variation. If the shadow label is also modeled as a stochastic oracle draw, then for fixed $h_i(y)$,
\[
    \mathrm{Var}\!\left[h_i(y)+\frac{\mathbb I[i\in U_t]}{\psi_{i,t}}(Z_i(y)-h_i(y))\right]
    =\frac{q_i(y)(1-q_i(y))}{\psi_{i,t}}
    +\left(\frac{1}{\psi_{i,t}}-1\right)(q_i(y)-h_i(y))^2.
\]
The first term is oracle variation under shadow-audit sampling; with deterministic decoding it is zero for point-mass transitions. The second term is residual response-surface error.

\subsection{Local Response-Surface Regularity and Base Error}
Rather than assuming a globally smooth LLM, define a local modulus of continuity for stratum $C_m$:
\[
    \omega_{m,t}(r)
    =
    \sup_{\substack{i,j\in C_m\\ d_z(\hat z_i^t,\hat z_j^t)\le r}}
    \|\boldsymbol q_{i,t}^{A}-\boldsymbol q_{j,t}^{A}\|_1 .
\]
A Lipschitz condition is the special case $\omega_{m,t}(r)\le L_{m,t}r$. For a non-prototype core agent $j\in C_m$, let $P_\kappa(j)$ be its queried-prototype support and let $w_{ji,t}$ be interpolation weights with $\sum_{i\in P_\kappa(j)}w_{ji,t}=1$. Define
\[
    n_{\mathrm{eff}}(j,t)=\frac{1}{\sum_{i\in P_\kappa(j)}w_{ji,t}^2}.
\]
When $\boldsymbol h_j^t$ is formed from one-hot prototype labels,
\[
    \mathbb E\|\boldsymbol h_j^t-\boldsymbol q_{j,t}^{A}\|_1
    \lesssim
    \sqrt{\frac{|\mathcal Y|}{n_{\mathrm{eff}}(j,t)}}
    +
    \sum_{i\in P_\kappa(j)}w_{ji,t}\omega_{m,t}(d_z(\hat z_i^t,\hat z_j^t)).
\]
Thus prototype label noise and local propagation bias jointly determine base response-surface error. For
\[
    \widehat{\boldsymbol p}_t^{\mathrm{base}}
    =\frac{1}{N}\sum_{i\in D_t}\boldsymbol Z_i^t+\frac{1}{N}\sum_{i\in F_t}\boldsymbol h_i^t,
\]
the same local modulus gives
\[
\begin{aligned}
    \mathbb E\|\widehat{\boldsymbol p}_t^{\mathrm{base}}-\bar{\boldsymbol p}_t^{A}\|_1
    \lesssim
    &\frac{1}{N}\sum_{j\in F_t}\left[
    \sqrt{\frac{|\mathcal Y|}{n_{\mathrm{eff}}(j,t)}}
    +\sum_{i\in P_\kappa(j)}w_{ji,t}\omega_{m(j),t}(d_z(\hat z_i^t,\hat z_j^t))
    \right] \\
    &+O\!\left(\frac{\sqrt{|\mathcal Y|\,|D_t|}}{N}\right).
\end{aligned}
\]
For APS, define the stratum-level residual second moment
\[
    R_{m,t}^{\mathrm{res}}
    =\frac{1}{|F_t\cap C_m|}\sum_{i\in F_t\cap C_m}
    \mathbb E\|\tilde{\boldsymbol Z}_i^t-\boldsymbol h_i^t\|_2^2.
\]
If $a_{m,t}$ audit units are allocated to stratum $m$ and the design is approximately equal-probability within that stratum, then
\[
    \mathbb E\|\widehat{\boldsymbol p}_t^{\mathrm{audit}}-\bar{\boldsymbol p}_t^{A}\|_1
    \lesssim
    O\!\left(\frac{\sqrt{|\mathcal Y|\,|D_t|}}{N}\right)
    +
    O\!\left(
    \sum_{m\in\mathcal M_{\mathrm{core}}}
    \frac{|F_t\cap C_m|}{N}
    \sqrt{\frac{|\mathcal Y|\,R_{m,t}^{\mathrm{res}}}{a_{m,t}}}
    \right).
\]
A poor response surface does not create design bias under positive shadow-audit inclusion, but it increases residual variance and therefore the shadow-audit budget required for precise correction.

\subsection{Tail-Protected Singleton Routing}
For a tail candidate $i$, the local-propagation bias that would arise from nearby core prototypes is
\[
    b_i^{\mathrm{smooth}}
    =\left\|\sum_{j\in P_\kappa(i)}w_{ij,t}\boldsymbol q_{j,t}^{A}-\boldsymbol q_{i,t}^{A}\right\|_1.
\]
If $i\in O_N$, APS queries the LLM directly rather than applying this smoothing operator, so this local-propagation bias is removed. This does not imply that tail agents are human-accurate or that all minority opinions are protected; it only states that selected isolated feature-space units are not overwritten by local prototype smoothing. The added cost is $O(TM_{\mathrm{out}}(N))=O(TN^\zeta)$ for $\zeta<1$.

\subsection{Multi-Round Decomposition}
Let
\[
    e_t^\star=\mathbb E\|\widehat{\boldsymbol p}_t^{\mathrm{audit}}-\bar{\boldsymbol p}_t^\star\|_1,
    \qquad
    c_t^{\mathrm{audit}}=\mathbb E\|\widehat{\boldsymbol p}_t^{\mathrm{audit}}-\bar{\boldsymbol p}_t^A\|_1.
\]
Then $e_t^\star\le c_t^{\mathrm{audit}}+\epsilon_{\mathrm{ctx},t}$. If context construction and the LLM transition have temporal sensitivity $a_t$, so that $\epsilon_{\mathrm{ctx},t}\le a_t e_{t-1}^\star$, then
\[
    e_t^\star\le a_t e_{t-1}^\star+c_t^{\mathrm{audit}},
    \qquad
    e_T^\star\le\sum_{t=1}^{T}\left(\prod_{\tau=t+1}^{T}a_\tau\right)c_t^{\mathrm{audit}}.
\]
This recurrence separates the current-round APS estimation terms from temporal context mismatch and explains why multi-round drift need not increase monotonically.

\subsection{Residual-Aware Allocation}
For core stratum $C_m$, the population analogue of the score in Eq.~\ref{eq:residual_risk_score} is
\[
    R_{m,t}=V_{m,t}^{\mathrm{res}}+\lambda_\rho L_{m,t}^2\rho_{m,t}^2+
    \lambda_e e_{m,t}^2+\lambda_r(1-r_{m,t})^2.
\]
Ignoring integer rounding and the minimum-one floor, the continuous relaxation
\[
    \min_{\{B_{m,t}\}}
    \sum_{m\in\mathcal M_{\mathrm{core}}}\frac{|C_m|^2}{N^2}\frac{R_{m,t}}{B_{m,t}}
    \quad\mathrm{s.t.}\quad
    \sum_m B_{m,t}\le B_t,
    \quad B_{m,t}\ge 0
\]
has Lagrangian first-order condition $B_{m,t}\propto |C_m|\sqrt{R_{m,t}}$. APS replaces $R_{m,t}$ with audit-estimated diagnostics,
\[
    \widehat R_{m,t}=\widehat V_{m,t}^{\mathrm{res}}+
    \lambda_\rho \widehat L_{m,t}^2\widehat\rho_{m,t}^2+
    \lambda_e \widehat e_{m,t}^2+
    \lambda_r(1-\widehat r_{m,t})^2,
    \qquad \tau>0,
\]
uses $\sqrt{\widehat R_{m,t}+\tau}$ as the allocation weight, and then applies rounding and the minimum-one constraint.

\subsection{Metric Connection and No-Free-Lunch}
The bounds above are stated in total variation or $L_1$ error. For finite $|\mathcal Y|=K$, $L_1$ control implies Jensen--Shannon control by standard entropy-continuity arguments: if $\|\widehat{\boldsymbol p}_t-\bar{\boldsymbol p}_t\|_1\le\varepsilon$, then
\[
    \mathrm{JSD}(\widehat{\boldsymbol p}_t,\bar{\boldsymbol p}_t)
    \le O\!\left(\varepsilon\log\frac{K}{\varepsilon}\right)
\]
for small $\varepsilon$. Exact match is stricter and is not the primary estimand. Finally, sublinear querying is impossible without structure: if no local regularity, tail-protected singleton routing, or positive shadow-audit coverage is assumed, then for binary outcomes one can construct two transition kernels that agree on all queried agents but assign opposite outcomes to all unqueried agents. Any $o(N)$-query method cannot distinguish them, yielding constant worst-case population error.

\section{Experimental Protocol and Reproducibility}
\label{app:reproducibility}

This appendix groups the experimental details into baseline implementations, reference construction and evaluation, call accounting, population construction, and runtime settings.

\subsection{Scaling-Table Baseline Selection}
\label{app:scaling_baseline_selection}
We include two scale-oriented baselines because they address different ways of reducing online LLM cost in social simulation. Light Society provides a large-scale learned-surrogate reference point \citep{guan2025earthscale}. TopoSim reports experiments up to 5,000 agents, but it proposes a topology-aware coordination strategy that explicitly reduces LLM calls by querying grouped representatives rather than all agents \citep{xu2026toposim}. We therefore adapt TopoSim as a grouped-coordinate baseline for the LLM-call saving setting in Table~\ref{tab:validation}.

\subsection{LS-Surrogate Implementation}
\label{app:ls_surrogate}
LS-Surrogate denotes the learned transition proxy adapted from Light Society. It replaces online LLM transition calls with neural-network inference during large-scale simulation. We train a compact transition classifier on reference-derived LLM labels; these labels are counted as method-side distillation calls in Table~\ref{tab:validation}. For the 10K--1M rows, the reported distillation budgets count labels across all eight rounds. For the 10M row, LS-Surrogate is trained from the completed 1M full-LLM rollout, corresponding to 8.0M labeled transition calls, and then rolled out to 10M agents by neural inference. The input is the 19-dimensional standardized WVS profile together with a learned round embedding for the eight scenario stages, and the output is a five-way opinion decision. Training uses an agent-level train/validation split to avoid placing the same respondent profile in both splits, class-weighted cross-entropy for the imbalanced option distribution, Adam optimization, early stopping, and the best validation checkpoint. During rollout, the trained classifier predicts every agent at every round and the top-1 class becomes the recurrent surrogate state. This baseline represents the learned-proxy family: it is efficient at inference time, but its transition rule is no longer the online LLM oracle and transfer to a new reference LLM, prompt format, population, opinion domain, or social-context representation generally requires constructing new LLM targets and retraining or fine-tuning the surrogate.

\subsection{TopoSim-Coord Implementation}
\label{app:toposim_coord}
TopoSim-Coord adapts the coordination-unit idea of TopoSim to our eight-round opinion-transition task. At each round, we construct a coordination feature for every agent by concatenating standardized profile features, a one-hot previous opinion state, the previous-round neighbor-opinion exposure vector, and graph degree, with weights $1.0$, $1.0$, $1.0$, and $0.25$, respectively. We then cluster agents into the scheduled number of coordination units with mini-batch $K$-means, choose the closest agent to each unit centroid as the representative, query the LLM once for each representative under the current scenario and recurrent context, and assign that representative decision to all agents in the same unit. Thus TopoSim-Coord uses one online LLM call per queried representative. The scale-first schedule is
\[
    K_{\mathrm{coord}}(N)=\left\lceil 2600\left(\frac{N}{10{,}000}\right)^{0.40}\right\rceil
\]
coordination units per round for $T=8$ rounds. Table~\ref{tab:toposim_coord_schedule} reports the resulting target schedule and the completed method-side calls used in Table~\ref{tab:validation}; the latter are the actual online representative queries recorded by the completed runs. The 10\,M JSD and exact-match entries are measured from the completed 10\,M TopoSim-Coord rollout against the corresponding 10\,M full-LLM reference. Unlike APS, TopoSim-Coord does not perform shadow-audit residual correction, tail-protected singleton routing, or residual-aware budget allocation.

\begin{table}[h]
\centering
\caption{TopoSim-Coord scale-first schedule used for Table~\ref{tab:validation}. Target calls are $8K_{\mathrm{coord}}(N)$; reported calls are the completed method-side online calls shown in Table~\ref{tab:validation}.}
\label{tab:toposim_coord_schedule}
\begin{adjustbox}{max width=\textwidth}
\begin{tabular}{lrrr}
\toprule
Scale & $K_{\mathrm{coord}}(N)$ per round & Target calls & Reported calls \\
\midrule
10K & 2,600 & 20.8K & 20.7K \\
100K & 6,531 & 52.2K & 52.2K \\
1M & 16,405 & 131.2K & 130.2K \\
10M & 41,208 & 329.7K & 329.7K \\
\bottomrule
\end{tabular}
\end{adjustbox}
\end{table}

\subsection{Same-Budget Approximation Baselines}
\label{app:same_budget_baselines}
Table~\ref{tab:same_budget} uses the completed 10K full-LLM trajectory as the evaluation reference, but reference labels are not used by the same-budget approximation baselines during simulation. All methods use the same 10K population, graph, scenario stages, option parser, model endpoint, deterministic decoding, and total budget of 15,500 online LLM calls over eight rounds. The budget is split as evenly as possible across rounds and then allocated across core strata in proportion to stratum size unless stated otherwise. Non-APS baselines use the same core stratification but do not use tail-protected singleton routing, shadow-audit residual correction, or residual-aware allocation.

Stratified sampling queries random agents within each core stratum and reports a stratum-weighted empirical distribution; recurrent hard states are filled by the sampled majority within each stratum. Cluster-based assignment uses the same sampled labels but assigns every agent in a stratum to the sampled majority label and reports the resulting hard-state distribution. Label propagation samples state-stratified anchors inside each stratum and assigns nonqueried agents by inverse-distance local propagation from same-stratum anchors, without shadow-audit residual correction. Core-set uses a static farthest-first representative order within each stratum; each round takes the next representatives under the matched budget and then applies the same local propagation rule.

The four stronger variants keep the same budget but change the selection or prediction mechanism. Stratified empirical estimates a full categorical distribution within each stratum from state-stratified queried labels and samples recurrent states from that distribution while retaining queried labels. Active selection weights both stratum budgets and within-stratum sampling toward high-dispersion, rare-previous-state, high-entropy, or high-change regions observed in previous rounds, then uses local propagation. Medoid clustering selects anchors by mini-batch $K$-means on profile features augmented with the previous state and chooses the nearest agent to each mini-cluster center before local propagation. Online surrogate trains a round-specific random-forest classifier on the queried labels using profile features plus previous-state one-hot features, predicts all agents, reports the mean predicted class probabilities, and uses the top class as the recurrent state.

\subsection{Reference Construction and Evaluation Protocol}
\label{app:evaluation_protocol}
For each reported population scale, the reference rollout is run independently of APS: every reference agent is updated by the corresponding reference LLM across all required rounds, and final-round targets come from the corresponding full-LLM trajectory. The 3K--100K rows use OpenRouter with glm-4.7-flash in one client process with four API keys; the 1M--10M rows use the official Zhipu API with glm-4-flash. The reference trajectory uses its own LLM-generated previous states and neighbor summaries. No APS state, propagated label, audit result, or budget decision enters the reference.

The reference, APS, LS-Surrogate, and TopoSim-Coord simulations use the same agent profiles, scenario text, option set, LLM backend, social-context construction protocol, and static graph at each reported scale. The graph is generated once for a scale and reused across the corresponding method and reference rollouts; only the evolving neighbor-state summaries differ because each rollout maintains its own recurrent states. APS, LS-Surrogate, and TopoSim-Coord are evaluated against the same corresponding full-LLM reference targets. JSD measures final-round distributional discrepancy against these targets, while exact match uses each method's recurrent hard top-1 state and measures aligned agreement with the reference label. The completed 10K full-trajectory check is the controlled setting used to inspect temporal distributional drift across all rounds. Table~\ref{tab:reference_manifest} reports the provider and client-configuration metadata for the main scaling table.

\begin{table}[h]
\centering
\caption{Reference construction manifest for Table~\ref{tab:validation}. Reference rollouts are independent of APS and use the provider/client configurations shown below.}
\label{tab:reference_manifest}
\begin{adjustbox}{max width=\textwidth}
\begin{tabular}{llll}
\toprule
Scale & Provider/model & Client configuration & Rounds \\
\midrule
3K & OpenRouter / glm-4.7-flash & 1 process, 4 keys, 100/key initial & 8 \\
10K & OpenRouter / glm-4.7-flash & 1 process, 4 keys, 100/key initial & 8 \\
100K & OpenRouter / glm-4.7-flash & 1 process, 4 keys, 100/key initial & 8 \\
1M & Zhipu / glm-4-flash & 2 processes, 30 keys, 60/key initial & 8 \\
10M & Zhipu / glm-4-flash & 5 processes, 50 keys, 60/key initial & 8 \\
\bottomrule
\end{tabular}
\end{adjustbox}
\end{table}

\subsection{Call Accounting}
\label{app:call_accounting}
The 10K and 100K APS rows in Table~\ref{tab:validation} are high-budget validation runs, whereas the 1M--10M APS rows follow the production schedule in Eq.~\ref{eq:sampling}. Table~\ref{tab:validation} reports method-side LLM calls and excludes the independent full-LLM reference rollout used for evaluation. APS call counts include core-prototype, singleton-tail, and shadow-audit queries. TopoSim-Coord call counts report grouped-representative LLM calls under its own scale-first schedule; the 10M JSD and exact-match entries are measured from the completed 10M TopoSim-Coord rollout. For LS-Surrogate, reported calls are the reference-derived LLM labels used to train the surrogate, counted across all eight rounds: 8.0K at 10K, 80.0K at 100K, and 800.0K at 1M. The 10M LS-Surrogate row uses the 1M full-LLM rollout as its distillation source, so it counts 8.0M labeled transition calls and has a $10.0\times$ reduction relative to the 80M-call brute-force 10M rollout. For every method, reduction is computed as the brute-force full-simulation call count at that scale divided by the reported LLM Calls entry. Bracketed values are 95\% confidence intervals where available.

\subsection{Provider-Specific Runtime and Cost Records}
\label{app:runtime_accounting}
Table~\ref{tab:small_runtime_cost} reports the 3K--100K full-reference and APS accounting corresponding to Table~\ref{tab:validation}. These runs use the OpenRouter chat-completions endpoint with glm-4.7-flash, one client process, four API keys, 100 initial concurrent requests per key, and checkpoint-based resume. Monetary cost is computed from the OpenRouter prices of \$0.06 per million input tokens and \$0.40 per million output tokens using the recorded input and output token usage for each run.

\begin{table}[h]
\centering
\caption{Runtime and cost accounting for the 3K--100K rows in Table~\ref{tab:validation}. Both full-reference simulation and APS use OpenRouter with glm-4.7-flash, one client process, four API keys, 100 initial concurrent requests per key, and checkpoint-based resume.}
\label{tab:small_runtime_cost}
\begin{tabular}{llrrr}
\toprule
Scale & Run & LLM calls & Wall-clock time & API cost \\
\midrule
3K & Full LLM reference & 24.0K & 26.5 min & \$1.79 \\
3K & APS & 8.9K & 9.8 min & \$0.69 \\
10K & Full LLM reference & 80.0K & 1.47 h & \$6.12 \\
10K & APS & 20.8K & 22.9 min & \$1.57 \\
100K & Full LLM reference & 800.0K & 14.70 h & \$60.14 \\
100K & APS & 172.0K & 3.16 h & \$12.94 \\
\bottomrule
\end{tabular}
\end{table}

Table~\ref{tab:large_runtime} reports the 1M--10M runtime accounting. These runs use the official Zhipu API with glm-4-flash. The 1M run uses 30 API keys across two client processes; the 10M run uses 50 API keys across five client processes. All large-scale runs start from 60 concurrent requests per key, use adaptive concurrency control for rate-limit backoff and retry scheduling, and support checkpoint-based resume. The APS times are end-to-end wall-clock records, including LLM calls, clustering, chunked feature processing, neighbor-summary construction, local propagation, audit aggregation, checkpoint flushing, and simulation bookkeeping.

\begin{table}[h]
\centering
\caption{Runtime accounting for the 1M--10M rows in Table~\ref{tab:validation}. Runs use the official Zhipu API with glm-4-flash; 1M uses 30 keys across two processes, while 10M uses 50 keys across five processes. All use 60 initial concurrent requests per key, adaptive concurrency control, and checkpoint-based resume.}
\label{tab:large_runtime}
\begin{tabular}{llrr}
\toprule
Scale & Run & LLM calls & Wall-clock time \\
\midrule
1M & Full LLM reference & 8.0M & 19.55 h \\
1M & APS & 83.4K & 2.80 h \\
10M & Full LLM reference & 80.0M & 9.56 d \\
10M & APS & 209.9K & 26.64 h \\
\bottomrule
\end{tabular}
\end{table}

\subsection{Confidence Intervals}
\label{app:confidence_intervals}
For reference-aligned evaluations, exact-match intervals use Wilson binomial intervals over aligned method-reference label pairs. JSD intervals use a paired nonparametric bootstrap over these completed method-reference pairs when labels are retained; when only category counts are retained, we use a conservative multinomial bootstrap around the two completed empirical marginals. These conditional agent-level reporting intervals shrink with the number of aligned evaluated agents. They do not mean that the reference itself is a sampled approximation, and they do not quantify model, prompt, graph, population-construction, run-to-run, or human behavioral uncertainty.

\subsection{Ablation Protocols}
\label{app:ablation_protocols}
The small-scale allocation ablation compares adaptive and fixed allocation at matched total online calls; the call count includes core-prototype, singleton-tail, and shadow-audit queries. Table~\ref{tab:audit_outlier_ablation} reports two component-specific 10K ablation blocks. Each block is evaluated against its corresponding full-LLM final-round reference frame, so matched-budget comparisons should be read within block rather than as cross-block absolute JSD comparisons. The shadow-audit residual-correction block uses the standard 10K full-LLM reference frame and reports both the lower-call no-correction setting at 9.3K calls and a matched-budget no-correction setting at 12.3K calls; the residual-correction setting also uses 12.3K calls over eight rounds. The tail-protected singleton routing block uses a tail-enriched 10K full-LLM reference frame containing all 330 routed feature-space tails plus 9,670 core agents, so the selected tail agents are represented in the reference-aligned evaluation. It likewise reports a lower-call no-routing setting at 9.4K calls, a matched-budget no-routing setting at 12.3K calls, and the routed setting at 12.3K calls.

\subsection{Population Features and Profile Construction}
\label{app:population_construction}
Agent profiles used in the reported simulations start from about 90K real respondent records in the multi-country World Values Survey \citep{wvs2022wave7}. We use 19 scenario-relevant survey fields as standardized profile features. Missing or negative special codes are set to zero for the standardized feature matrix, and all selected fields are standardized with \texttt{StandardScaler} before stratification, tail scoring, and local interpolation. The selected fields are:
\begin{center}
\begin{adjustbox}{max width=\textwidth}
\small
\begin{tabular}{llll}
\toprule
WVS code & Field & WVS code & Field \\
\midrule
\texttt{X003R} & age group & \texttt{Q262} & gender \\
\texttt{Q273} & education level & \texttt{Q274} & income level \\
\texttt{Q275} & marital status & \texttt{N\_REGION\_WVS} & region \\
\texttt{G\_TOWNSIZE} & city size & \texttt{Q3} & health status \\
\texttt{Q46} & social trust & \texttt{Q55} & political interest \\
\texttt{Q57} & social activity level & \texttt{Y003} & secular values \\
\texttt{I\_AUTHORITY} & authority respect & \texttt{I\_NATIONALISM} & nationalism \\
\texttt{DEFIANCE} & defiance & \texttt{SCEPTICISM} & scepticism \\
\texttt{AUTONOMY} & autonomy & \texttt{EQUALITY} & equality belief \\
\texttt{I\_DEVOUT} & religiosity &  &  \\
\bottomrule
\end{tabular}
\end{adjustbox}
\end{center}
The survey variables are used as profile features for prompt construction, stratification, tail scoring, and local interpolation. For target populations no larger than the empirical WVS pool, we sample respondent records directly. For larger target populations, we expand the pool by empirical-seed perturbation rather than by independent Gaussian noise. We first draw seed respondents with replacement within coarse country/region, age, gender, education, and income cells when those cells are available; otherwise we back off to the closest nonempty marginal cell. Categorical variables such as gender, marital status, and region are kept on their legal WVS support and are either copied from the seed respondent or redrawn from the empirical conditional distribution of the same cell. Ordinal variables such as education, income, health, trust, and political interest are perturbed by a discrete local kernel that favors the seed category and adjacent categories, then clipped to the valid WVS code range. Continuous value indices such as authority respect, nationalism, defiance, scepticism, autonomy, equality belief, secular values, and religiosity are perturbed in standardized space using a shrinkage local covariance estimated from nearest empirical respondents in the same coarse cell; values are then clipped to empirical quantile bounds before standardization.

This mixed-type expansion keeps every generated profile anchored to a real WVS respondent, respects categorical and ordinal support, and preserves local correlations more faithfully than applying independent Gaussian noise to each feature. It is still a simulation input, not a validated demographic reconstruction of any real population. We therefore interpret the large-scale populations as computational stress tests over WVS-derived profile variation rather than as claims about true national or global population composition.

\subsection{Social Graph and Recurrent Context}
\label{app:graph_context}
For each population scale, we construct one static sparse Watts--Strogatz-style graph before round 1 and reuse it across the corresponding method and reference rollouts. The default context degree is $k_G=10$: each agent sees ten previous-round neighbor states in its prompt. We start from a ring lattice and rewire right-side edges with probability $p=0.1$, using the experiment seed. The graph is stored as an \texttt{int32} neighbor array of shape $(N,k_G)$, so memory is $O(Nk_G)$; at 10M agents and $k_G=10$, the array requires about 0.37 GiB. The graph is not conditioned on APS strata or WVS groups and has no explicit homophily parameter. Rewired and ring edges can cross strata, so social context can transmit previous-round opinions across clusters. The graph is fixed across rounds and across matched method-reference rollouts; only the neighbor-state summaries change because each rollout maintains its own recurrent states.

\subsection{Stratification, Response Surface, and Tails}
\label{app:stratification_response_surface_tails}
Core strata follow the schedule in Eq.~\ref{eq:cluster_schedule}. The production implementation uses mini-batch or streaming-compatible clustering and never forms an $N\times N$ distance matrix. The 10K backend checks additionally evaluate Gaussian mixture and BIRCH alternatives under the same evaluation protocol. For non-prototype core agents, the local response surface is computed only within the same core stratum. If $P_\kappa(j)$ is the set of the $\kappa$ nearest queried prototypes to agent $j$ in standardized feature space, we use inverse-distance weights
\[
w_{ji,t}=\frac{(d(x_j,x_i)+10^{-6})^{-1}}
{\sum_{\ell\in P_\kappa(j)}(d(x_j,x_\ell)+10^{-6})^{-1}},
\qquad \kappa=5.
\]
The value $\kappa=5$ was fixed before the reported experiments after pilot checks: smaller supports made prototype-label noise visible, while larger supports increased smoothing over isolated, heterogeneous, or high-curvature feature-space regions. Tail-protected singleton routing selects $M_{\mathrm{out}}(N)$ singleton-tail agents before core clustering. In the reported implementation, the tail score is the robust standardized distance from the coordinate-wise median,
\[
s_i=\left\|\frac{x_i-\mathrm{median}(X)}{\mathrm{MAD}(X)}\right\|_2,
\]
with small MAD entries floored for stability; the top-scoring agents are directly queried every round. Exact Euclidean distances are used at validation scale. At larger scales, feature generation, clustering, and within-stratum interpolation are processed in chunks, and only queried prototypes must be indexed for nearest-support search.

\subsection{LLM Calls, Parsing, and Logging}
\label{app:llm_calls_logging}
All reported APS call counts include core-prototype, singleton-tail, and shadow-audit queries. The 3K--100K GLM runs use the OpenRouter-compatible chat-completions endpoint \texttt{https://openrouter.ai/api/v1} with model identifier glm-4.7-flash, one client process, four API keys, 100 initial concurrent requests per key, and checkpoint-based resume. The 1M--10M GLM runs use the official Zhipu endpoint \texttt{https://open.bigmodel.cn/api/paas/v4} with model identifier glm-4-flash; 1M uses 30 official API keys across two client processes, while 10M uses 50 official API keys across five client processes. Large-scale runs start from 60 concurrent requests per key and use adaptive concurrency control to reduce provider errors plus checkpoint-based resume. The six-backend robustness runs use the provider interface and exact model identifiers shown in Table~\ref{tab:model_generalization}. Runs reported in this submission were executed in April--May 2026 with deterministic decoding, temperature 0, top-p 1.0, maximum 500 tokens, 120-second timeout, and up to three API retries. Provider APIs do not expose immutable checkpoint hashes, so each result file records the model identifier, base URL, decoding parameters, seeds, and run date. Responses are parsed into the five allowed options; failed or unparseable responses are retried, and included runs have zero unresolved unparseable decisions after repair. Initial parse failures and API retry counts are recorded in the run artifacts by round and batch. The exact prompt template and scenario text are reproduced below.

\subsection{Shadow-Audit Inclusion Probabilities and Seeds}
\label{app:audit_inclusion_seeds}
For a core stratum $C_m$ with correction frame $F_t\cap C_m$ and shadow-audit allocation $a_{m,t}$, equal-probability shadow auditing gives $\psi_{i,t}=a_{m,t}/|F_t\cap C_m|$ before clipping at one. When state-stratified shadow-audit sampling is used, the recorded inclusion probability is the product of the predicted-state cell selection probability and the uniform probability within that cell. The APS residual formula first produces an unprojected signed estimate whose entries sum to one by construction and support the design-unbiasedness result. Before JSD and table reporting, we apply a simplex projection: negative entries are clipped to zero and the vector is renormalized to sum to one; if all entries are zero after clipping, a uniform fallback is used. We record both the pre-projection vector and the reported projected vector in artifacts when shadow-audit residual correction is enabled. The artifacts record round number, stratum id, prototype ids, singleton-tail ids, shadow-audit ids, allocation counts, and random seeds, which is sufficient to reconstruct $\psi_{i,t}$ for the APS estimator. Unless otherwise noted, population, graph, clustering, reference rollout, and LLM sampling use seed 42; the two-seed sampling ablation uses seeds 42 and 43.

\subsection{Scale Resources}
\label{app:scale_resources}
The 10M run stores recurrent states as \texttt{int8}, graph neighbors as \texttt{int32}, and profile features either as \texttt{float32} chunks or a materialized 19-feature matrix when memory allows. A materialized 10M $\times$ 19 \texttt{float32} feature matrix is about 0.71 GiB, and the 10-neighbor graph is about 0.37 GiB. The reported 10M APS pipeline ran on a Linux CPU server with two AMD EPYC 7543 processors, 128 hardware threads, and 2 TiB RAM, using chunked feature loading and vectorized neighbor summaries; Table~\ref{tab:large_runtime} gives the corresponding wall-clock runtime. APS itself does not require a GPU, although LS-Surrogate uses PyTorch acceleration when available.

\section{Prompt Templates and Scenario Text}
\label{app:prompts}

This appendix reports the prompt structure and scenario text used in the experiments. The main scaling experiments and all main-scenario ablations use the same 8-round subway crisis scenario and the same decision template. The three scenario-generalization experiments use the same template with different scenario text and option sets.

\subsection{Generic LLM Decision Template}

For every queried core prototype, singleton-tail agent, shadow-audit agent, or full-reference agent, the LLM receives an agent profile, optional interaction context from local neighbors, the current scenario stage, and a mutually exclusive option set. The prompt text is shown in a shaded box to distinguish it from the paper text.

\begin{promptbox}[title=Generic LLM Decision Template]
You are an ordinary online user following this event.

Agent profile: {formatted demographic, value, trust, and social-context features}

Previous attitude: {previous-round state, when available}

Local social context: {summarized neighbor attitudes, when available}

Current event {round index}: {scenario stage text}

Please choose one option from the following list:
1. {option_1}
2. {option_2}
...
K. {option_K}

Decision requirements: consider your own background and standpoint; consider all previous events rather than only the current message; your view may change as the event develops.

Return JSON only:
{"decision": "1", "reasoning": "brief reason grounded in the agent profile"}
\end{promptbox}

The system instruction is: \emph{You are a rational decision maker; make a judgment based on your background, experience, and all known information.} Failed or unparseable responses are retried; unresolved responses are treated as failures rather than randomly mapped to an option.

\subsection{Main and Ablation Scenario: Subway Chemical Attack}

The main scaling experiments and all main-scenario ablations use the following five options throughout all eight rounds:

\begin{promptbox}[title=Main Scenario Option Set]
1. Fully support the current government response and trust that the government will handle the crisis properly.
2. Generally support the government response, but ask for more transparency and public participation.
3. Stay neutral and wait for more information before deciding.
4. Have substantial doubts about the government response and ask for reassessment.
5. Completely distrust the government's handling capacity and demand an independent third-party investigation.
\end{promptbox}

The eight scenario stages are:

\begin{promptbox}[title=Subway Chemical Attack Scenario Stages]
1. Subway gas attack. During the morning rush hour, a suspected toxic-gas leak occurs on a subway platform. Twelve people die, more than 200 are hospitalized, and more than 30 are in critical condition. The subway line is shut down and panic spreads. Social media circulates competing explanations: a terrorist attack, a chemical leak, or an accident caused by aging subway infrastructure.
2. Official classification. Police classify the event as an organized terrorist attack and identify a suspect described as an outsider with extremist views. Anonymous claims suggest the suspect may instead be a dismissed subway maintenance worker seeking revenge. Officials announce a special task force and ask residents not to spread rumors.
3. Citywide security escalation. The government launches a top-level counter-terror response: subway passengers must arrive 30 minutes early for security checks, temporary checkpoints appear at malls and schools, and migrant workers must re-register identity information. Some residents ask whether the registration policy is discriminatory. Officials frame the measures as temporary public-safety protections.
4. Economic shock. One week later, subway ridership falls by 60%, businesses along the line lose half their revenue, and some foreign firms consider reducing investment. The city announces subsidies, but large chain firms receive more than small shops. Delivery workers become essential while their labor protections remain weak. Officials state that recovery funds will prioritize small firms and individual businesses.
5. Leaked internal report. A leaked subway safety audit claims that engineers warned about serious ventilation defects three months before the attack, but management postponed repairs because of budget constraints. The subway company calls the document fake but provides no counter-evidence. Victims' families demand a thorough investigation. Officials say they are verifying the document.
6. International reaction. Several countries issue travel warnings and some flights are suspended. An international human-rights organization calls for protection of citizens' rights to information and safety. The foreign ministry rejects politicization of a security incident. Domestic opinion divides between supporting the official stance and welcoming international supervision.
7. Accountability storm. The provincial discipline commission investigates the subway group chairman. Public attention shifts to why only one person is being investigated, as media summarize seven major safety accidents in five years and a pattern of temporary dismissal followed by return to office. Families' lawsuits are delayed pending the criminal investigation. Officials promise a thorough disciplinary and judicial process.
8. Reconstruction dispute. After one month of shutdown, two camps emerge: an efficiency camp wants rapid reopening to stop economic losses, while a safety camp demands complete safety upgrades even if reopening is delayed by three months. Residents face a trade-off between commute convenience and safety doubts. The city announces a hearing, but the selection of hearing representatives is itself questioned.
\end{promptbox}

\subsection{Scenario-Generalization Prompts}

Each generalization experiment uses 10K agents, six rounds, the generic template above, and one of the following scenario-specific option sets.

\paragraph{Polarized AI Surveillance Policy.}
\begin{promptbox}[title=Polarized AI Surveillance Policy Prompt]
Options: (1) strongly support city-wide deployment; (2) support limited deployment with independent oversight; (3) undecided or wait-and-see; (4) oppose deployment unless major safeguards are added; (5) strongly oppose any deployment.

Stages: proposal of AI cameras in transit hubs and busy streets for violent-crime reduction and missing-person cases; pilot safety evidence showing faster emergency response; privacy warnings and false-alert examples from civil-rights groups; vendor metadata leak and suspension of vendor access; proposed independent audit board, deletion rules, and complaint channels; final city-council vote on full deployment, limited pilot, postponement, or cancellation.
\end{promptbox}

\paragraph{Minority-Sensitive Digital Healthcare Reform.}
\begin{promptbox}[title=Minority-Sensitive Digital Healthcare Reform Prompt]
Options: (1) strongly support immediate adoption; (2) support adoption with offline alternatives; (3) neutral or need more information; (4) oppose unless vulnerable groups are protected; (5) strongly oppose because it excludes vulnerable groups.

Stages: hospital announcement of a unified app for appointments, AI triage, and queue management; early efficiency gains for working-age users; complaints from older adults, rural residents, and low-income patients who have difficulty using the app or lack smartphones; a widely shared missed-care case after failed online identity verification; offline windows, family authorization, hotline booking, and staff assistance are added; final citywide rollout decision after the protection package.
\end{promptbox}

\paragraph{Coordination and Contagion in Evacuation.}
\begin{promptbox}[title=Coordination and Contagion in Evacuation Prompt]
Options: (1) evacuate immediately; (2) prepare to evacuate if neighbors do; (3) wait for official confirmation; (4) stay home unless danger becomes visible; (5) refuse evacuation.

Stages: severe-storm watch with possible flooding within 48 hours; voluntary advisory and mixed evacuation behavior across communities; visible neighbor signals as some leave, some board windows, and others dismiss warnings; rising water and stronger recommendation for exposed neighborhoods; conflicting rumors about shelter crowding versus volunteer reports that transport remains available; final window before transport may stop and late rescue becomes difficult.
\end{promptbox}

\section{Robustness Protocol Notes}
\label{app:robustness_results}

The robustness results are reported in the main text in Table~\ref{tab:robustness_combined} and Figure~\ref{fig:generalization}. Scenario tests use three 10K-agent, six-round tasks with a fixed 20\% prototype rate and full-LLM references on the same agents. The LLM backend rows repeat these scenario checks across six completed backends; one attempted backend returned no parseable option decisions and is excluded. The stratification backend rows replace the partitioner while fixing the population, prompt, LLM backend, and full-LLM reference.

\section{Ethics and Social Impact}
\label{app:ethics}

APS is a tool for approximating a computational object: the population transition induced by a specified LLM, prompt, population representation, and scenario. It should not be interpreted as measuring real public opinion, true policy preferences, or human crisis behavior. LLM-generated decisions can reflect model biases, prompt framing, training-data artifacts, and limitations of the input profile data. We therefore report uncertainty intervals where applicable, separate computational fidelity from behavioral validity, and avoid using full-LLM references as evidence of human truth.

Large-scale attitude simulation has dual-use risks. The same machinery that can support stress tests of public communication or policy scenarios could also be misused for opinion manipulation, propaganda optimization, micro-targeting, or group profiling. We do not endorse deployment for real political persuasion, voter targeting, crisis messaging optimization, or decisions that materially affect individuals or protected groups without independent human-subject review, domain oversight, and validation against appropriate empirical data. Outputs should be used only as exploratory model diagnostics unless a separate governance process establishes legitimacy, consent, and accountability.

Minority and vulnerable groups require particular care. APS includes tail-protected singleton routing to reduce smoothing of isolated feature-space regions, but this is not a guarantee that minority viewpoints or lived experiences are correctly represented. Results for small or high-curvature groups should be reported with uncertainty, shadow-audit disagreement, and subgroup diagnostics; unsupported subgroup claims should be withheld. If shadow-audit diagnostics reveal high residual error for a group, the appropriate response is to increase validation and disclose the limitation, not to propagate the simulated result as a population fact.

The WVS empirical seed pool and perturbation procedure also carry limitations. Survey coverage, nonresponse, measurement choices, cultural framing, and expansion choices can distort the represented agents. These biases can be amplified by LLM prompts and local response surfaces. Any release or downstream use should document the data source, profile construction, prompt text, model version, uncertainty estimates, known failure modes, and prohibited uses.

\end{document}